\documentstyle[11pt]{article}

\newcommand{\sect}[1]{\setcounter{equation}{0}\section{#1}}
\newcommand{\subsect}[1]{\subsection{#1}}
\renewcommand{\theequation}{\thesection.\arabic{equation}}
\newcommand{\vs}[1]{\rule[- #1 mm]{0mm}{#1 mm}}
\newcommand{\hs}[1]{\hspace{#1 mm}}
\newcommand{\lbl}[1]{\label{eq:#1}}
\newcommand{\rf}[1]{(\ref{eq:#1})}
\newcommand{\nn}{\nonumber\\}
\newcommand{\eq}{\vs{2}\large\begin{equation}}
\newcommand{\en}{\\[2mm]\end{equation}\normalsize}
\newcommand{\cen}{\\[2mm]\]\normalsize}
\newcommand{\ceq}{\vs{2}\large\[}
\newcommand{\bea}{\large\begin{eqnarray}}
\newcommand{\ena}{\end{eqnarray}\normalsize}
\newcommand{\nnbea}{\large\begin{eqnarray*}}
\newcommand{\nnena}{\end{eqnarray*}\normalsize}
\newcommand{\leqa}{\lefteqn}

\newcommand{\grt}[1]{\mbox{\LARGE #1}}
\newcommand{\prt}{\partial}
\newcommand{\bz}{\bar{z}}
\newcommand{\prtb}{\bar{\partial}}
\newcommand{\lra}{\longrightarrow}
\newcommand{\tld}{\widetilde}

\newcommand{\mbf}[1]{\mbox{\boldmath $#1$}}

\newcommand{\muz}{\mu^z_{\bz}}
\newcommand{\mub}{\bar{\mu}}

\newcommand{\mmb}{\mu \bar{\mu}}
\newcommand{\lambdab}{\bar{\lambda}}
\newcommand{\jb}{\bar{\jmath}}
\newcommand{\bc}{\bar{c}}
\newcommand{\bC}{\bar{C}}
\newcommand{\bZ}{\bar{Z}}
\newcommand{\zbz}{(z,\bar{z})}
\newcommand{\ZBZ}{(Z,\bar{Z})}
\newcommand{\ZBZp}{(Z',\bar{Z'})}

\newcommand{\paZ}{\partial_Z}

\newcommand{\pabZ}{\partial_{\bar{Z}}}
\newcommand{\cdotd}{(c \! \cdot \! \prt)}
\newcommand{\DER}[1]{\frac{D}{D #1 \zbz}}

\newcommand{\J}[2]{{{\cal J}{}^{}}^{#1}_{#2,\star} \zbz}
\newcommand{\bigJ}[2]{{{\cal J}{}^{}}^{#1}_{#2,\star} \ZBZ}
\newcommand{\jnat}[2]{{{\cal J}^\natural}^{#1}_{#2,\star} \zbz}

\newcommand{\jhat}[2]{\widehat{{\cal J}{}}^{#1}_{#2,\star} \zbz}
\newcommand{\jhhat}[2]{\widehat{\widehat{\cal J}{}}^{#1}_{#2,\star}
\zbz}

\newcommand{\jnatz}[2]{{{\cal J}^\natural}^{#1}_{#2z,\star} \zbz}
\newcommand{\jnatbz}[2]{{{\cal J}^\natural}^{#1}_{#2\bar{z},\star}
\zbz}
\newcommand{\jnatzbz}[2]{{{\cal
J}^\natural}^{#1}_{#2z\bar{z},\star} \zbz}
\newcommand{\jnatzero}[2]{{{\cal J}^{\natural,0}}^{#1}_{#2,\star}
\zbz}

\newcommand{\DDEL}[2]{{\Delta{}^{}}^{#1}_{#2} \zbz}
\newcommand{\delnat}[2]{{\Delta^\natural}^{#1}_{#2} \zbz}
\newcommand{\delhat}[2]{\widehat{\Delta{}}^{#1}_{#2,\star} \zbz}

\newcommand{\NP}[1]{Nucl.\ Phys.\ {\bf #1}}

\newcommand{\PL}[1]{Phys.\ Lett.\ {\bf #1}}

\newcommand{\PRev}[1]{Phys.\ Rev.\ {\bf #1}}

\font\twelve=cmbx10 at 15pt
\font\ten=cmbx10 at 12pt

\begin{document}
\renewcommand{\thefootnote}{\fnsymbol{footnote}}
\begin{titlepage}

\begin{center}
{\ten Centre de Physique Th\'eorique\footnote{Unit\'e Propre de
Recherche 7061} - CNRS - Luminy, Case 907}

{\ten F-13288 Marseille Cedex 9 - France }

\vspace{2 cm}
{\twelve DIFFEOMORPHISM COHOMOLOGY}

{\twelve IN BELTRAMI PARAMETRIZATION II : THE 1-FORMS}

\vskip 1truecm
\setcounter{footnote}{0}
\renewcommand{\thefootnote}{\arabic{footnote}}

{\bf Giuseppe BANDELLONI\footnote{e-mail:
BEPPE@GENOVA.INFN.IT}$^,$\footnote{Istituto di Fisica della
Facolt\`a di Ingegneria, Universit\`a
degli Studi di Genova,  P.le Kennedy, I-16129 GENOVA, Italy  and
Istituto Nazionale di Fisica Nucleare, INFN, Sezione di Genova,
via Dodecaneso 33, I-16146 GENOVA, Italy}
 and Serge LAZZARINI
\footnote{also: Universit\'e d'Aix-Marseille II}$^,$\footnote{e-mail:
SEL@CPT.UNIV-MRS.FR}
}
\vspace{1.5 cm}

{\bf Abstract}
\end{center}

We study the 1-form diffeomorphism cohomologies
within a local conformal   Lagrangian Field Theory model
built on a two dimensional Riemann surface with no boundary.
We consider the case of scalar matter fields  and the complex
structure is
parametrized by Beltrami differential.
The analysis is first performed at
the Classical level, and then we improve  the quantum
extension, introducing the current in the Lagrangian dynamics,
coupled to external source fields.
We show that the anomalies which spoil the current conservations
take origin from the holomorphy region of the external fields, and
only the
differential spin 1 and 2 currents (as well their c.c) could be
anomalous.
\vskip 1truecm
\centerline{To be published in Journ. Math. Phys.}

\vskip 2truecm
\noindent P.A.C.S. 11.10.Gh/03.70

\bigskip
\noindent September 1994

\noindent CPT-94/P.3072

\bigskip

\noindent anonymous ftp or gopher: cpt.univ-mrs.fr

\end{titlepage}


\sect{Introduction}

\indent

The most transparent and useful formulation for a field Theory
surely is
the one in which the locality is manifest.

The interest for this approach has been raised from the physical
relevance
of the local symmetries, such as gauge invariance for particle
Physics and
the diffeomorphism role in String and Gravitational models.

Now it appears in pratice that, for system of physical interest in
which
a local symmetry is realized, pointing out locality is a good
appeling to investigate the deep meaning of the symmetry through
the study of those  local objects which, due to their invariance,
have global properties.

This is the reason why it has been necessary to introduce in the
literature
the so called  descent equations , and, plunged in the B.R.S.
approach, they
gave outstanding  results in hunting anomalies, vertex operators,
and so on.

In this scheme it has been necessary to provide the local objects
as a form
graduation, in relation to their invariance.

In a recent paper \cite{theone} we have used this strategy to study
the
anomalies \cite{BRS} and the vertex operators in two dimensional
diff-invariant models
in Beltrami parametrization \cite{BB,B,BBS,LS,KLT,KLS}, seen as
local density quantities
with form graduation
equal to two, since they are to be integrated in two dimensional
manifolds.

Indeed Beltrami parametrization is a good chance for studying the
chiral
resolution in conformal blocks \cite{BPZ} and holomorpic
factorization
\cite{BK,BB,B,LS,BBS}, since the complex structure parametrization
is realized in
an automatic way.

In this paper we want to investigate the objects with form
graduation equal
to one, that is currents, which play an essential role in the
symmetry
realization and in the understanding of the links between the
Lagrangian
and the Hamiltonian approaches.
In \cite{witten1,witten2} this analysis has been used in the
Hamiltonian
spirit, so its improvement within a Lagrangian
framework is required.

Indeed the  $C^{\infty}$ smooth holomorphic functions

\bea
\biggl(z,\bz\biggr) \lra \biggl(Z(z,\bz),\bZ(z,\bz)\biggr)
\lbl{diffchart}
\ena
will individuate a set $\ZBZ$ of coordinates and
$\zbz$ will be a  reference frame of holomorphic coordinates on a
Riemann
surface.

Under this action, the set $\ZBZ$ can be transformed  under a
reparametrization
of $\zbz$; so the invariance under the diffeomorphism:
\bea
\biggl(Z(z,\bz),\bZ(z,\bz)\biggr) \lra
\biggl(Z'(z,\bz),\bZ'(z,\bz)\biggr)
\lbl{diffchart1}
\ena
can individuate models of relevant physical interest.

It is well known that infinitesimal action of \rf{diffchart1}
realizes a nilpotent operator s :
\eq \begin{array}{rl}
sZ &=\ (c^z\prt + c^{\bz}\prtb)Z \nn
sc^z &=\ (c^z\prt + c^{\bz}\prtb)c^z \nn
s^2 &=\ 0 \ .
\end{array}
\lbl{SS}
\en
where the ghost fields $c^z\zbz,c^{\bz}\zbz$ carry a
$Q_{\Phi\Pi}$ charge equal to one.

So the diffeomorphism will change the $\ZBZ$ coordinates by means
the action
on the $\zbz$ ones.

The Beltrami parametrization:
\bea
dZ  = \lambda (dz + \mu d\bz)\nn
\lambda = \prt Z \nn
\mu = \frac{\prtb Z}{\prt Z}\nn [3mm]
\lbl{dZ}
\ena
with the compatibility condition:
\bea
(\prtb - \mu \prt) \lambda = \lambda \prt \mu
\lbl{intfac}
\ena
is particularly attractive since in this approach conformal
rescaling is
seen as a diffeomorphism $\ZBZ\lra\ZBZp$ of the surface into itself
with the same
$\mu$.
Indeed when the Beltrami parameters are taken as constant the
reparametrization
operation is represented by complex analytic transition functions.
 Furthermore equivalence class of analytic atlases will identify a
complex
structure, and a conformal classe of 2-dimensional metrics.
That is identifying
conformal invariant models on a Riemaniann manifold needs a
carefull
description to factorize this equivalence arbitrariness.
However, not  all diffeomorphisms of the surface into itself amount
to just
a conformal rescaling.
The intrinsic geometry of the surface is determined  by the metric
tensor (which is a coordinate-free object), and other changes  in
the metric
produce confomally inequivalent surfaces.

Furthermore in a Lagrangian Field Theory model the Beltrami
differentials
are  the appropriate sources of the energy momentum tensor
components,
whose short distance products will define the algebraic
construction \cite{BPZ}
of conformal current algebra.

Our purpose is here to characterize in a cohomological way all the
diffeomorphism conserved currents, first at a Classical level an
then
to extend their properties (first of all their conservation) to the
Quantum one.

Being the Action an invariant $(1,1)$ tensor, we shall suppose that
in a
conformal invariant theory the matter fields are realized on the
Riemaniann
manifold by  local  tensor fields $\Phi_{j\jb} \ZBZ dZ^j
d\bZ^{\jb}$ of
weight ${(j,\jb)}$ invariant under change of holomorphic charts
\bea
\Phi_{j\jb} {\ZBZ}^{\alpha} dZ^{j}_{\alpha} d\bZ^{\jb}_{\alpha}
=\Phi_{j\jb} {\ZBZ}^{\beta} dZ^{j}_{\beta} d\bZ^{\jb}_{\beta}\nn
\lbl{invtensor}
\ena
It is possible also to define, via the diffeomorphism action
restricted to
dilatations, the geometric dimensions defined as:

\eq
{\rm Dim} = N(\downarrow) - N(\uparrow) + Q_{\Phi\Pi}\ .
\lbl{id}
\en

The B.R.S. realization of the  infinitesimal
diff-variations then reads:

\eq \begin{array}{rl}
sZ &=\ \gamma^Z \equiv \lambda^Z_z(c^z + \muz c^{\bz}) \\[2mm]
s \Phi_{j\jb} \ZBZ &=\ (\gamma^Z \paZ +
\gamma^{\bZ}\pabZ)\Phi_{j\jb}\ZBZ\\[2mm]
s \gamma^Z &=\ s \gamma^{\bZ}=0\\[2mm]
s^2 &=\ 0 \ .
\end{array}
\lbl{S}
\en

and its complex conjugates.

Going to little "z" indices it writes:

\eq
\begin{array}{rl}
s Z &=\ \lambda (c + \mu \bc)\equiv\gamma \\[2mm]
s \lambda &=\ \cdotd \lambda + \lambda (\prt c + \mu \prt
\bc)\equiv\prt(
\lambda( c+\mu \bc)) \\[2mm]
s \mu &=\ \cdotd \mu - \mu (\prt c + \mu\prt\bc) + \prtb c +
\mu\prtb\bc \\[2mm]
sc &=\ \cdotd c \\[2mm]
s \phi_{j,\jb} &=\ \cdotd \phi_{j,\jb} + \grt{(} j(\prt c +
\mu\prt\bc)
+ \jb (\prtb\bc + \mub\prtb c) \grt{)} \phi_{j,\jb}\ ,
\end{array}
\lbl{s}
\en
with of course the complex conjugate expressions.

The matter fields are parametrized as:

\bea
\Phi_{j\jb} \ZBZ dZ^j d\bZ^{\jb} &=& \phi_{j\jb} \zbz (dz + \muz
\zbz d\bz)^j
(d\bz + \mub \zbz dz)^{\jb}\nn
&&\mbox{}\lbl{rescal}\\[-2mm]
\ena
with:
\bea
\Phi_{j\jb} \ZBZ\ &=& \frac{\phi_{j\jb} \zbz}{\lambda^j\zbz
\lambdab^{\jb}\zbz}.
\nn
\ena

The previous variations \rf{s} will define a B.R.S. local operator
$\delta$ such that
$\delta^2=0$ acting on the space of the previous fields and their
derivatives
considered as independent monomials coordinates as in Ref
\cite{dixon}

So, even if the $\ZBZ$ frame will describe the model, the use of
little
"z" coordinates is particularly useful (as remarked in \cite
{theone,beppe}, since the the derivative operator can be defined,
in the
above mentioned scheme,by means of the $\delta$ operator and the
"little" c
ghosts as we shall see in the following).

 Let  $Q_{\star}$ be the physical charges in each tensorial
sectors,
such that the  $\quad{\star}$   label will sum up covariant and
controvariant "big"
indices.
These charges will derive from currents as:
\eq
Q_{\star} = \int \grt{(} {\cal J}_{Z,\star} dZ + {\cal
J}_{\bZ,\star} d\bZ \grt{)} =
\int \bigJ{0}{1}
\lbl{bigQ}
\en
so that the form degree is with respect the "big" indices.

The diffeomorphism invariance will assure that it will be a
counterpart in the "little"
indices:

\eq
Q_{\star} = \int \grt{(} {\cal J}_{z,\star} dz + {\cal
J}_{\bz,\star} d\bz \grt{)} =
\int \J{0}{1}
\lbl{littleQ}
\en

The aim of this paper is to study the diff-invariant charges, that
is
the quantities $Q_{\star}$ which verify:

\eq
\delta Q_{\star} = 0.
\en

It is well known that such a symmetry has to require, at least at
the
quantum level, conserved currents,
which is a "local" constraint well defined with respect to a
reference frame;
but the diff-invariance \rf{diffchart} puts on the same footing a
large class
of systems of coordinates: so it may be interesting to investigate
how
the symmetry, realized in a local way, will generalize at each
chart
the currents conservation.
We shall find that the two dimensional character of the theory,
if it is defined on manifolds without boundary, requires, for the
existence of the charge, the  holomorphic factorization of currents
in the
$Z(z,\bz)$ (or its c.c) variable.
This fact will have many important consequences that we shall
investigate in
this paper. First of all the fact that all the local currents will
have
definite covariance properties, that is their will be
$(n+1,0)$ (or their c.c) true tensors.

Furthermore our aim will be to extend at the Quantum level all the
properties found, established at the classical one.
For this reason we have to put the currents inside an invariant
Action by
coupling them to external fields, and we have to study the
perturbative
renormalization of the model.
We shall show that only the spin 1 and 2 currents will spoil
conservation  at the quantum level, while all the other ones will
maintain all the classical symmetries.
The paper is organized as follows:

In Section 2 we shall solve the 1-forms descent equation deriving from
 the  diff-mod {\bf{d}} invariance  and in particular we shall show
that diff-current conservations can be derived from the diff-cohomology,
so they cannot depend on the particular choice of coordinates:
we shall establish these relations  in the "true" $\ZBZ$
coordinates
as well as in the reference frame $\zbz$.We shall show that the
charge existence
condition on a Riemann surface with no boundary, implies
holomorphic
constraints for the currents.

In Section 3  we shall  briefly recover the previous results in a
Lagrangian two dimensional dynamics, which forces a two-forms
analysis.
This artillery allows a perturbative quantum extension of the
diff-invariance
in order  to find the possible obstructions to both the current
covariance
and the current conservation.

An Appendix is devoted to some computational details using the
spectral sequences method \cite{dixon}\cite{beppe}

\sect{The 1-forms in the $\delta$-cohomology}

\indent

In this section we want to analyze the descent equation of 1-forms,
already
done in \cite{witten1}\cite{witten2}, but following the spirit of
\cite{theone}.

To be more accurate we shall first relate  the diff-mod {\bf{d}}
cohomology to the local
unintegrated  functions, by solving the ordinary
differential action in term of the B.R.S. operator for
diffeomorphism.

Furthermore we shall show that all the uncharged elements which are
solutions
of the 1-forms descent equations will indentify conserved currents:
 more
exactly, the diffeomorphism cohomology alone will specialize
current
conservation both on the "little" and "big" coordinates:

This fact has an important consequence for the two-dimensional
character of
the theory: the current conservation condition will admit inversion
formula,
and, on a manifold without boundary, the holomorphic factorization
of currents
will be obtained.

We stress that this  is, in our framework, a classical level
analysis
which support
a particular importance for proving the stability properties of
theory;
the quantum extension will need more accurancy.

\subsection{The consistency equations and the current conservation}

\indent

We shall start from those objects (defined in the general
reference frame $\zbz$)

\eq
Q_{\star} = \int \grt{(} {\cal J}_{z,\star} dz + {\cal
J}_{\bz,\star} d\bz \grt{)} =
\int \J{0}{1}
\lbl{littleQ1}
\en
(and the form degree is relative to the "little" indices)

which are elements of the $\delta$ cohomology:
\eq
\delta Q_{\star} = 0\\;
Q_{\star}\neq \delta \hat{Q}_{\star}
\en

In terms of local quantities the cocycle equation will imply:
\eq
\begin{array}{r}
\delta \J{0}{1} + d \J{1}{0} = 0 \\[1mm]
\end{array}
\lbl{descent0}
\en
where  $\delta$ operator acts in the space of local unintegrated
functions as
described in \rf{s}; its full complete description will be found in
the
Appendix\rf{delta00}.

The previous equation will characterize $\J{0}{1}$ as an element of
the
diff-mod  $\bf{d}$ cohomology .

 From \rf{descent0} we derive the expressions
of the well-known descent equations:
\eq
\begin{array}{r}
\delta \J{0}{1} + d \J{1}{0} = 0 \\[1mm]
\delta \J{1}{0} = 0
\end{array}
\lbl{descent}
\en
so that the bottom current writes
\eq
\J{1}{0} = \jnat{1}{0} + \delta \jhat{0}{0}
\en
where $\jnat{1}{0}$ is an element of the cohomology of $\delta$ in
the space
of the unintegrated functions.

Writing the differential operator in term of the $\delta$  as in
\cite{beppe,theone}

\bea
\prt=\biggl\{\delta,\DER{c}\biggr\}\nn
\prtb=\biggl\{\delta,\DER{\bc}\biggr\}\nn
\lbl{diff1}
\ena
(we remark that it is only true in the "little" coordinates
reference frame)
one gets by a direct substitution in eqs.\rf{descent}

\nnbea \leqa{
\delta \J{0}{1} + \left( dz \grt{\{} \delta , \DER{c} \grt{\}} +
d\bz \grt{\{} \delta , \DER{\bc} \grt{\}} \right) (\jnat{1}{0} +
\delta \jhat{0}{0}) =}
 \\[2mm]
&& = \delta \left[ \J{0}{1} + \left( dz \DER{c} +
d\bz \DER{\bc} \right) \jnat{1}{0}
+ d \jhat{0}{0} \right] = 0
\nnena
which is solved by the 1-form
\eq
\J{0}{1} = \jnat{0}{1}
- \left( dz \DER{c} + d\bz \DER{\bc} \right) \jnat{1}{0}
- d \jhat{0}{0} + \delta \jhat{-1}{1}
\lbl{coho}
\en

This is the fundamental formula which relates the  elements of the
diff-mod $\bf{d}$ cohomology to the  elements of  of $\delta$ one.

It  will be very useful to calculate this cohomological space: this
will
be done below. Let us however recall the most important result
coming from this calculation and show their consequences.

First of all, denoting by $N_{z}(\downarrow)$ and $N_{z}(\uparrow)$
the lower
and upper "little" indices counting operators respectively, we
shall show in
the Appendix  that:

\bea
(N_{z}(\downarrow)-N_{z}(\uparrow)) \jnat{0}{1} =
(N_{\bz}(\downarrow)-N_{\bz}(\uparrow)) \jnat{0}{1} = 0
\ena

which as 1-form implies:
\eq
\jnat{0}{1} \equiv 0
\en
so \rf{coho} reduces to:

\eq
\J{0}{1} = - \left( dz \DER{c} + d\bz \DER{\bc} \right) \jnat{1}{0}
- d \jhat{0}{0} + \delta \jhat{-1}{1}
\lbl{cohocont1}
\en

and the dimensions of the currents are given by the Z(big!) index
content.
 \eq
{\rm Dim} = N_{Z}(\downarrow) - N_{Z}(\uparrow) \ .
\lbl{id1}
\en
where $N_{Z}(\downarrow)$ and $ N_{Z}(\uparrow)$ are, as can be
easily understood,
the counting operators of the "big" lower and upper indices
respectively.

Then \rf{cohocont1} tell us that the
current conservation is a direct  consequence of the diffeomorphism
invariance.

Indeed, as pointed out in \cite{witten1}
applying d on eq. \rf{descent0}, we get:
\eq
\begin{array}{r}
d\delta \J{0}{1} = 0 \\[1mm]
\end{array}
\lbl{descent01}
\en

that is, since $\biggl[ d, \delta \biggr]=0$,
in terms of the $\delta$-cohomology functions, we get:

\eq
\begin{array}{r}
d \J{0}{1} =\J{0,\natural}{2}+\delta\widehat{\J{-1}{1}}  \\[1mm]
\end{array}
\lbl{descent02}
\en

we want here to show  that the $\J{0}{2}^{\natural}$ term
of the r.h.s. is zero.

 From the very definition we have:

\bea \leqa{
d\J{0}{1} =
- \left(\prtb \DER{c}\jnat{1}{0}dz\wedge d\bz +
\prt \DER{\bc}\jnat{1}{0}
d\bz\wedge dz \right)+\delta d \jhat{-1}{1}}\nn
&&=\left(-\prtb \DER{c}\jnat{1}{0}+
\prt \DER{\bc}\jnat{1}{0}\right)
d\bz\wedge dz+\delta d \jhat{-1}{1} \nn
\lbl{diffj1}
\ena
on the one hand we have from \rf{diff1}

\bea \leqa{}
d\J{0}{1} =
\left(-\biggl\{\delta,\DER{\bc}\biggr\}
\DER{c}\jnat{1}{0}\right.\nn
\left.+\biggl\{\delta,\DER{c}\biggr\} \DER{\bc}\jnat{1}{0}\right)
d\bz\wedge dz+\delta d \jhat{-1}{1} \nn
\lbl{diffj2}
\ena

On the other hand, by using directly, in \rf{diffj1}:

\bea
\biggl[\prt, \DER{\bc}\biggr]=\biggl[\prtb, \DER{c}\biggr]=0
\lbl{comm}
\ena
we get

\bea \leqa{}
d\J{0}{1} =- {\left( \DER{c}\prtb\jnat{1}{0}dz\wedge d\bz +
 \DER{\bc}\prt\jnat{1}{0}d\bz\wedge dz \right)}\nn
+\delta d \jhat{-1}{1}-\left(
\DER{c}\biggl\{\delta,\DER{\bc}\biggr\}
\jnat{1}{0}\right.\nn
\left.+ \DER{\bc}\biggl\{\delta,\DER{c}\biggr\}\jnat{1}{0}\right)
d\bz\wedge dz+\delta d \jhat{-1}{1} \nn
\lbl{diffj3}
\ena

Comparison of eqs now \rf{diffj2} and \rf{diffj3}:

\eq
d\J{0}{1} = - \delta\left( \DER{c}\DER{\bc}\jnat{1}{0}d\bz\wedge dz
- d \jhat{-1}{1}\right)\\[1mm]
\lbl{diffj20}
\en
that is, the $\J{0}{2}^{\natural}$ obstruction   term, which a
priori occurred
in \rf{descent02}
does not appear.

Moreover the term $\DER{c}\DER{\bc}\jnat{1}{0}d\bz\wedge dz$ can
contribute
to the  $\Phi,\Pi$ uncharged sector only if $\jnat{1}{0}$ will
contain
negative charged fields:so,  since we remove the anti-ghosts
fields,
after imposing their equations of motion through the gauge fixing,
we can
assume the only $\Phi,\Pi$ charged negative fields are only those
which occur
in the Lagrangian coupled to the B.R.S. transformation.

We shall show in the Appendix that in the  $\delta$ cohomology
space no
negative charged field  can appear, hence:

\eq
d\J{0}{1}
=\delta d{\jhhat{-1}{1}}\\[1mm]
\lbl{diffj30}
\en
but recalling that $\J{0}{1}$ is a representative of an equivalence
class, and is
defined modulo arbitrary $\delta$ contributions, the current
conservation is
realized only for  elements $\J{0}{1}-\delta{\jhhat{-1}{1}}$ which
will
define "locally" the  conserved current ${\tilde\J{0}{1}}$, such
that:

\eq
d{\tilde\J{0}{1}}\equiv
d\left(\J{0}{1}-\delta{\jhhat{-1}{1}}\right) = 0
\lbl{diffj4}
\en

\subsect{The local $\delta$ cohomology}

\indent

As pointed out in the formula \rf{cohocont1}, we have shown that
the
elements of the
diff-mod {\bf{d}} cohomology can be easily derived from the ones of
the
$\delta$ cohomology in the space of local functions.
The aim of this part is to solve:
\eq
\delta \jnat{r}{0}=0
\lbl{locoho}
\en
where the upper index $r$ will label the $\Phi,\Pi$ charge and the
lower index
the form degree respectively. The $\Phi,\Pi$ charge sector we are
interested
in, is the one with $r=1$

If we decompose the cohomology spaces into their underivated ghost
content:
\eq
\jnat{r}{0}=\jnatzero{r}{0} + c \jnatz{r-1}{0} +\bc \jnatbz{r-1}{0}
+c \bc
\jnatzbz{r-2}{0}
\lbl{decomp1}
\en
where $\jnatzero{r}{0},\jnatz{r-1}{0},\jnatbz{r-1}{0},
\jnatzbz{r-2}{0}$ do not contain  underivated ghosts,
the condition \rf{locoho} will imply the following system:

\bea\leqa{
\widehat{\delta}\jnatzero{r}{0}=0}
\\[1mm]
&& \biggl(-\widehat{\delta}+\prt c \biggr)\jnatz{r-1}{0}+\prt
\bc\jnatbz{r-1}
{0}+\prt\jnatzero{r}{0}=0
\\[1mm]
&& \biggl(-\widehat{\delta}+\prtb  \bc \biggr)\jnatbz{r-1}{0}
+\prtb c\jnatz{r-1}{0}
+\prtb\jnatzero{r}{0}=0 \\[1mm]
&&\biggl(\widehat{\delta}-\prt c-\prtb\bc \biggr)\jnatzbz{r-2}{0}
-\prtb \jnatz{r-1}{0}+\prt\jnatbz{r-1}{0}=0 \\[1mm]
\lbl{decomp2}\nonumber
\ena
where
\eq
\widehat{\delta}\equiv \delta- c\prt-\bc\prtb
\lbl{deltatilde}
\en
hence:
\bea\leqa{
\widehat{\delta}\jnatzero{r}{0}=0}
\lbl{1eq}
\\[2mm]
&& \biggl(-\widehat{\delta}+(\prt-\mub\prtb) c \biggr)(
\jnatz{r-1}{0}-\mub\jnatbz{r-1}{0})
+(\prt-\mub\prtb)\jnatzero{r}{0}=0
\\[2mm]
&& \biggl(-\widehat{\delta}+(\prtb -\mu\prt)  \bc
\biggr)(\jnatbz{r-1}{0}-
\mu\jnatz{r-1}{0})
+(\prtb-\mu\prt)\jnatzero{r}{0}=0 \\[1mm]
&&\biggl(\widehat{\delta}-\prt c-\prtb\bc \biggr)\jnatzbz{r-2}{0}
-\prtb \jnatz{r-1}{0}+\prt\jnatbz{r-1}{0}=0 \\[1mm]
\lbl{decomp3}\nonumber
\ena

In the Appendix we shall show that in a general Lagrangian model,
in which,
for the sake of simplicity the matter fields are taken to be
scalar,
and the only
$\Phi,\Pi$ negative charged fields are the external ones coupled to

the B.R.S. variations (and the gauge terms are taken away by
solving
$\widehat{\delta}^2=0$) the $\widehat{\delta}$ cohomology space
does not
contain negative charged fields, so for
$r\neq 0$ the solution of \rf{1eq} is $\widehat{\delta}$ trivial:

\eq
\jnatzero{r}{0}=\widehat{\delta}{\jhat{r-1}{0}}
\en

Then, using:
\eq
\biggl[\widehat{\delta},\prt-\mub\prtb\biggr]=\biggl(
(\prt-\mub\prtb)c
\biggr) (\prt-\mub\prtb)
\en
we get
\bea\leqa{}
&& \biggl(-\widehat{\delta}+(\prt-\mub\prtb) c \biggr)\biggl(
\jnatz{r-1}{0}-\mub\jnatbz{r-1}{0}
+(\prt-\mub\prtb){\jhat{r-1}{0}}\biggr)=0
\\[2mm]
&& \biggl(-\widehat{\delta}+(\prtb -\mu\prt)  \bc
\biggr)\biggl(\jnatbz{r-1}{0}-
\mu\jnatz{r-1}{0}
+(\prtb-\mu\prt)\bar{\jhat{r-1}{0}}\biggr)=0 \\[1mm]
\lbl{decomp4}\nonumber
\ena

So, if we define:
\eq
\jnatz{r-1}{0}-\mub\jnatbz{r-1}{0}+(\prt-\mub\prtb){\jhat{r-1}{0}}\equiv
\lambda {\cal J}^{r-1}_{Z,0,\star}\ZBZ(1-\mu\mub)
\en
we get
\eq
\widehat{\delta}{\cal J}^{r-1}_{Z,0,\star}\ZBZ=0
\en
which is solved, according to the results given in the Appendix,by:
\eq
{\cal J}^{r-1}_{Z,0,\star}\ZBZ={\cal
J}^{\natural,r-1}_{Z,0,\star}\ZBZ+\widehat
{\delta}\widehat{{\cal J}}^{r-2}_{Z,0,\star}\ZBZ
\en
where ${\cal J}^{\natural,r-1}_{Z,0,\star}\ZBZ$ and $\bar{\cal
J}^{\natural
,r-1}_{\bZ,0,\star}\ZBZ$ are elements of the $\widehat{\delta}$
cohomology
space.

Next defining:
\eq
\jnatz{r-1}{0}=\prt\jhat{r-1}{0}+\lambda({\cal
J}^{\natural,r-1}_{Z,0,\star}
\ZBZ+
\widehat{\delta}\widehat{
{\cal J}}^{r-2}_{Z,0,\star}\ZBZ)+\mub\lambdab(\bar{{\cal
J}}^{\natural,r-1}_
{\bZ,0,\star}
\ZBZ+\widehat{\delta}\widehat{
\bar{{\cal J}}}^{r-2}_{\bZ,0,\star}\ZBZ)
\en

\eq
\jnatbz{r-1}{0}=\prtb\jhat{r-1}{0}+\lambdab(\bar{{\cal
J}}^{\natural,r-1}_
{\bZ,0,\star
}\ZBZ+
\widehat{\delta}\widehat{
\bar{{\cal J}}}^{r-2}_{\bZ,0,\star}\ZBZ)+\mu\lambda({{\cal
J}}^{\natural,r-1}_
{Z,0,\star}
\ZBZ+\widehat{\delta}\widehat{
{{\cal J}}}^{r-2}_{Z,0,\star}\ZBZ)
\en
we have
\bea\leqa{}
&&\biggl(\widehat{\delta}-\prt c-\prtb\bc \biggr)\jnatzbz{r-2}{0}
-\prtb \jnatz{r-1}{0}+\prt\jnatbz{r-1}{0} \nn
&&\equiv\biggl(\widehat{\delta}-\prt c-\prtb\bc
\biggr)\jnatzbz{r-2}{0}
-\lambda\lambdab(1-\mu\mub)\biggl(\partial_{\bZ}({{\cal
J}}^{\natural,r-1}_{Z,0,
\star}
\ZBZ+\widehat{\delta}\widehat{
{{\cal J}}}^{r-2}_{Z,0,\star}\ZBZ)\nn
&&-\partial_{Z}(\bar{{\cal J}}^{\natural,r-1}_{\bZ,0,\star
}\ZBZ+
\widehat{\delta}\widehat{
\bar{\cal J}}^{r-2}_{\bZ,0,\star}\ZBZ)\biggr)=0\nn
\lbl{decomp5}
\ena

By introducing:

\eq
\jnatzbz{r-2}{0}=\lambda\lambdab(1-\mu\mub)
\Lambda^{r-2}_{Z,\bZ,\star}\ZBZ
\lbl{definl}
\en
we obtain:

\eq
\widehat{\delta}\biggl(\Lambda^{r-2}_{Z,\bZ,\star}\zbz-\partial_{\bZ}
\widehat{
\bar{\cal J}}^{r-2}_{\bZ,0,\star}\ZBZ+\partial_{\bZ}
\widehat{
{\cal J}}^{r-2}_{\bZ,0,\star}\ZBZ\biggr)-\partial_{\bZ}
{\cal J}^{\natural,r-1}_{Z,0,\star}\ZBZ
+\partial_{Z}\bar{\cal J}^{\natural,r-1}_{\bZ,0,\star}\ZBZ=0
\lbl{decomp6}
\en

Since $\partial_{\bZ}$ and $\partial_{Z}$ commute with
$\widehat{\delta}$, then
$\partial_{\bZ}{\cal J}^{\natural,r-1}_{Z,0,\star}\ZBZ$ and
$\partial_{Z}\bar{\cal J}^
{\natural,r-1}_{\bZ,0,\star}\ZBZ$ are elements of the same space,
so
 the only possibility to verify \rf{decomp6} is:

\eq
\widehat{\delta}\biggl(\Lambda^{r-2}_{Z,\bZ,\star}\ZBZ-\partial_{\bZ}
\widehat{
\bar{\cal J}}^{r-2}_{\bZ,0,\star}\ZBZ+\partial_{\bZ}
\widehat{
{\cal J}}^{r-2}_{\bZ,0,\star}\ZBZ\biggr)=0
\lbl{decomp61}
\en

\eq
-\partial_{\bZ}
{\cal J}^{\natural,r-1}_{Z,0,\star}\ZBZ
+\partial_{Z}\bar{\cal J}^{\natural,r-1}_{\bZ,0,\star}\ZBZ=0
\lbl{decomp62}
\en

Furthermore, since the $\widehat{\delta}$ cohomology does not
depend on the
external negative charged fields,solving \rf{decomp61} gives:
\eq
\biggl(\Lambda^{r-2}_{Z,\bZ,\star}\ZBZ-\partial_{Z}
\widehat{
\bar{\cal J}}^{r-2}_{\bZ,0,\star}\ZBZ+\partial_{\bZ}
\widehat{
{\cal J}}^{r-2}_{Z,0,\star}\ZBZ\biggr)=\widehat{\delta}\widehat{
{\Lambda}}^{r-3}_{Z,\bZ,\star}\ZBZ
\lbl{decomp63}
\en
so the final decomposition writes:
\bea
\jnat{r}{0}&=&\gamma{\cal
J}^{\natural,r-1}_{Z,0,\star}\ZBZ+\bar{\gamma}
\bar{\cal J}^{\natural,r-1}_{\bZ,0,\star}\ZBZ \nn
&+&{\delta}\biggl(\gamma\bar{\gamma}
\widehat{{\Lambda}}^{r-3}_{Z,\bZ,\star}\ZBZ \nn
&+&\bar{\gamma}\widehat{\bar{\cal
J}}^{r-2}_{\bZ,0,\star}\ZBZ+{\gamma}
\widehat{{\cal J}}^{r-2}_{Z,0,\star}\ZBZ \nn
&+&\jhat{r-1}{0}
\biggr) \nn
\lbl{decompfin1}
\ena
We emphasize that, in the "big" coordinates, the current
${\cal J}^{\natural,r-1}_{Z,0,\star}\ZBZ$ will always  transform as
a scalar
quantity, despite of its tensorial $\star$ content, that is:

\eq
\delta{\cal
J}^{\natural,r-1}_{Z,0,\star}\ZBZ=(\gamma\paZ+\bar{\gamma}\pabZ)
{\cal J}^{\natural,r-1}_{Z,0,\star}\ZBZ
\en

On the other hand, introducing the local quantities:

\eq
{\cal{S}}^{\natural,r-1}_{z,0,\star}\zbz\equiv\lambda{\cal
J}^{\natural,r-1}_
{Z,0,\star}\ZBZ
\\[1mm]
\lbl{slocal}
\en
\eq
\bar{\cal{S}}^{\natural,r-1}_{\bz,0,\star}\zbz\equiv\bar{\lambda}
\bar{\cal J}^{\natural,r-1}_{\bZ,0,\star}\ZBZ
\\[1mm]
\lbl{sbarlocal}
\en
they will transform in the"little" c ghosts as:

\bea
\delta{\cal{S}}^{\natural,r-1}_{z,0,\star}\zbz &=&
(c\prt+\bc\prtb){\cal{S}}^{\natural,r-1}_{z,0,\star}\zbz\nn
&+&(\prt c+\mu\prt\bc){\cal{S}}^{\natural,r-1}_{z,0,\star}\zbz
\lbl{Svar11}
\ena
so that it is a (1,0) tensor, while, going to the $C\zbz$ ghosts,
we get:

\bea
\delta{\cal{S}}^{\natural,r-1}_{z,0,\star}\zbz &=& \prt\biggl(
C{\cal{S}}^{\natural,r-1}_{z,0,\star}\zbz\biggr)\nn
&+&\biggl(\prtb
{\cal{S}}^{\natural,r-1}_{z,0,\star}\zbz-\prt(\mu{\cal{S}}^{\natural,r-1}_
{z,0,\star}\zbz)\biggr)\frac{(\bC\zbz -\mub\zbz C\zbz)}
{(1-\mu\mub)}\nn
\lbl{Svar1}
\ena

Similarly for its c.c. counterpart:

\bea
\delta\bar{\cal{S}}^{\natural,r-1}_{\bz,0,\star}\zbz &=&
(c\prt+\bc\prtb)\bar{\cal{S}}^{\natural,r-1}_{\bz,0,\star}\zbz\nn
&+&(\prtb\bc+\mub\prtb
c)\bar{\cal{S}}^{\natural,r-1}_{\bz,0,\star}\zbz
\lbl{Svar22}
\ena

\bea
\delta\bar{\cal{S}}^{\natural,r-1}_{\bz,0,\star}\zbz &=&
\prtb\biggl(\bC\bar{\cal{S}}^{\natural,r-1}_{\bz,0,\star}\zbz\biggr)\nn
&+&\biggl(\prt
\bar{\cal{S}}^{\natural,r-1}_{\bz,0,\star}\zbz-\prtb
(\mub\bar{\cal{S}}^{\natural,r-1}_
{\bz,0,\star}\zbz)\biggr)\frac{(C\zbz -\mu\zbz\bC\zbz)}
{(1-\mu\mub)}\nn
\lbl{Svar2}
\ena
(it is a (0,1) tensor )

So we can rewrite \rf{decompfin1} as:

\bea
\jnat{r}{0} &=& C\zbz
{\cal{S}}^{\natural,r-1}_{z,0,\star}\zbz
+\bC\zbz
\bar{\cal{S}}^{\natural,r-1}_{\bz,0,\star}\zbz \nn
 &+&{\delta}\biggl(C\zbz\bC\zbz\lambda
\bar{\lambda}
\widehat{{\Lambda}}^{r-3}_{Z,\bZ,\star}\ZBZ \nn
&+&\bC\zbz
\bar{\lambda}
\widehat{
\bar{\cal J}}^{r-2}_{\bZ,0,\star}\ZBZ  \nn
&+&C\zbz{\lambda}
\widehat{
{\cal J}}^{r-2}_{Z,0,\star}\ZBZ
+\jhat{r-1}{0}\biggr) \nn
\lbl{decompfin11}
\ena

A complete description in terms of the c's can be achieved
introducing:

\eq
\jnatz{r-1}{0}={\cal{S}}^{\natural,r-1}_{z,0,\star}\zbz+\mub
\bar{\cal{S}}^{\natural,r-1}_{\bz,0,\star}\zbz
\en
\eq
\jnatbz{r-1}{0}=
\bar{\cal{S}}^{\natural,r-1}_{\bz,0,\star}\zbz+
\mu{\cal{S}}^{\natural,r-1}_{z,0,\star}\zbz
\en
and \rf{decompfin11} then reads:

\bea
\jnat{r}{0}
&=&c\zbz
\jnatz{r-1}{0}
+\bc\zbz
\jnatbz{r-1}{0} \nn
 &+&{\delta}\biggl(c\zbz\bc\zbz(1-\mu\mub)\lambda
\bar{\lambda}
\widehat{{\Lambda}}^{r-3}_{Z,\bZ,\star}\ZBZ \nn
&+&(\bc\zbz+\mub\zbz c\zbz)
\bar{\lambda}
\widehat{
\bar{\cal J}}^{r-2}_{\bZ,0,\star}\ZBZ  \nn
&+&(c\zbz +\mu\zbz \bc\zbz){\lambda\zbz}
\widehat{
{\cal J}}^{r-2}_{Z,0,\star}\ZBZ
+\jhat{r-1}{0}\biggr) \nn
\lbl{decompfin12}
\ena

Therefore our final result for the currents \rf{cohocont1}, by
specializing
r=1, writes:
\eq
\J{0}{1} =  - \left( dz \DER{c} + d\bz \DER{\bc} \right)
\jnat{1}{0}
+ d{\widehat\J{0}{0}}      +\delta{\widehat\J{-1}{1}}\nn
\en
\eq
={\cal{S}}^{\natural,0}_{z,0,\star}\zbz(dz+\mu d\bz)+
\bar{\cal{S}}^{\natural,0}_{\bz,0,\star}\zbz(d\bz+\mub dz)
+ d{\widehat\J{0}{0}}      +\delta{\widehat\J{-1}{1}}\nn
\en

\eq
\equiv
\jnatz{0}{0} dz+
\jnatbz{0}{0} d\bz
+ d{\widehat{\J{0}{0}}}      +\delta{\widehat{\J{-1}{1}}}
\\[1mm]
\lbl{result}
\en
or, by covariance:
\eq
\J{0}{1}=
{\cal J}^{\natural,0}_{Z,0,\star}\ZBZ dZ+
\bar{\cal J}^{\natural,0}_{\bZ,0,\star}\ZBZ d\bZ
+ d {\widehat{\cal J}}^{0}_{Z,0,\star}\ZBZ
+\delta{\widehat{\cal J}}^{-1
}_{Z,0,\star}\ZBZ
\\[1mm]
\lbl{result1}
\en

Finally we want to remark that $\delta\jnat{r}{0} = 0$ implies, by
using
the decomposition \rf{decompfin1} :

\bea
\gamma\bar{\gamma}
\biggl(-\partial_{\bZ}
{\cal J}^{\natural,r-1}_{Z,0,\star}\zbz
+\partial_{Z}\bar{\cal J}^{\natural,r-1}_{\bZ,0,\star}\zbz
\biggr)\nn
&=&\frac{C\zbz\bC\zbz}{(1-\mu\zbz\mub\zbz)}\biggl(\prtb
{\cal{S}}^{\natural,r-1}_{z,0,\star}\zbz-\prt(\mu{\cal{S}}^
{\natural,r-1}_{z,0,\star}\zbz)\nn
&-&\prt\bar{\cal{S}}^{\natural,r-1}_{\bz,0,\star}\zbz+
\prtb(\mub\bar{\cal{S}}^{\natural,r-1}_{\bz,0,\star}\zbz)
\biggr)\nn
&=&{c\zbz\bc\zbz}
\biggl(-\prtb \jnatz{r-1}{0}+\prt\jnatbz{r-1}{0}\biggr)
= 0
\ena
the last equality tells us that
that the diffeomorphism invariance will imply the current
conservation both
in the "big" index current ${\cal
J}^{\natural,r-1}_{Z,0,\star}\zbz$
and in the "little" index one $\jnatz{r-1}{0}$, that is
$d{\J{0}{1}}=0$
as before.

\subsection{The charge definition and the holomorphic factorization
of currents}

\indent

The two dimensional character of the theory has an important
consequence, since
the current conservation condition can be inverted.
In fact from the condition \rf{decomp62}:

\bea
\biggl(-\partial_{\bZ}
{\cal J}^{\natural,r-1}_{Z,0,\star}\ZBZ
+\partial_{Z}\bar{\cal J}^{\natural,r-1}_{\bZ,0,\star}\ZBZ
\biggr)
= 0
\ena
it will formally follow:
\bea
{\cal J}^{\natural,r-1}_{Z,0,\star}\ZBZ=
{(\partial_{\bZ})}^{-1}\partial_{Z}\bar{\cal J}^
{\natural,r-1}_{\bZ,0,\star}\ZBZ
\ena
where the inverse operator ${(\partial_{\bZ})}^{-1}$ can be
defined, only
in two dimension,  using the Cauchy theorem:

\bea
{\cal J}^{\natural,r-1}_{Z,0,\star}\ZBZ=
\int_{\bf{C}} dW\wedge d\bar{W} \frac{\partial_{W}\bar{\cal J}^
{\natural,r-1}_{\bar{W},0,\star}(W,\bar{W})}{W-Z}\nn
=\int_{\bf{C}} dW\wedge d\bar{W}\partial_{W}\biggl( \frac{\bar{\cal
J}^
{\natural,r-1}_{\bar{W},0,\star}(W,\bar{W})}{W-Z}\biggr)
-\int_{\bf{C}} dW\wedge d\bar{W}\partial_{W}\biggl( \frac{1}{W-Z}
\biggr)\bar{\cal J}^
{\natural,r-1}_{\bar{W},0,\star}(W,\bar{W})\nn
=\partial_{Z} \int_{\bf{C}} dW\wedge d\bar{W} \frac{\bar{\cal J}^
{\natural,r-1}_{\bar{W},0,\star}(W,\bar{W})}{W-Z}+
\int_{\prt W}\frac{\bar{\cal J}^
{\natural,r-1}_{\bar{W},0,\star}(W,\bar{W})}{W-Z}\nn
\ena
and on a manifold without boundary  it reduces to:

\bea
{\cal J}^{\natural,r-1}_{Z,0,\star}\ZBZ=
{(\partial_{Z})}{\Omega}^
{\natural,r-1}_{0,\star}\ZBZ
\ena
so no charge $Q_{\star}$ can be obtained unless:

\bea
\partial_{\bZ}
{\cal J}^{\natural,r-1}_{Z,0,\star}\ZBZ=0\nn
\partial_{Z}\bar{\cal J}^{\natural,r-1}_{\bZ,0,\star}\ZBZ   = 0
\ena
hence, holomorphicity in Z (big!) is a necessary condition in order

to get charges.

In the z-frame, by \rf{slocal} and \rf{sbarlocal},
the existence of charges is assured only if we require the
supplementary
conditions:

\bea
\prtb
{\cal{S}}^{\natural,r-1}_{z,0,\star}\zbz=\prt(\mu{\cal{S}}^
{\natural,r-1}_{z,0,\star}\zbz) \nn
\prt
\bar{\cal{S}}^{\natural,r-1}_{z,0,\star}\zbz=\prtb
(\mub\bar{\cal{S}}^{\natural,r-1}_
{z,0,\star}\zbz)\nn
\lbl{holocurrent}
\ena

Accordingly, the $\delta$ variations  \rf{Svar1} \rf{Svar2} read:
\bea
\delta{\cal{S}}^{\natural,r-1}_{z,0,\star}\zbz=\prt\biggl(
C\zbz{\cal{S}}^{\natural,r-1}_{z,0,\star}\zbz\biggr)\nn
\ena
\bea
\delta\bar{\cal{S}}^{\natural,r-1}_{\bz,0,\star}\zbz=\prtb\biggl(\bC\zbz
\bar{\cal{S}}^{\natural,r-1}_{\bz,0,\star}\zbz\biggr)
\ena
we remark that if we define the currents
${\cal{S}}^{\natural,r-1}_{z,0}\zbz$
and $\bar{\cal{S}}^{\natural,r-1}_{\bz,0,\star}\zbz$ from their
previous BRS variations,
then   the conditions \rf{holocurrent} will derive from the
required
conditions \rf{diff1}:
\bea
\prt\bar{\cal{S}}^{\natural,r-1}_{\bz,0,\star}\zbz &=&
\biggl\{\delta,\DER{c}\biggr\}
\bar{\cal{S}}^{\natural,r-1}_{\bz,0,\star}\zbz\nn
&=&\prtb(\mub\bar{\cal{S}}^{\natural,r-1}_
{z,0,\star}\zbz)\\[3mm]
\prtb{\cal{S}}^{\natural,r-1}_{z,0,\star}\zbz
&=&\biggl\{\delta,\DER{\bc}\biggr\}
{\cal{S}}^{\natural,r-1}_{z,0,\star}\zbz\nn
&=&\prt(\mu{\cal{S}}^
{\natural,r-1}_{z,0,\star}\zbz)\nn
\lbl{diffcond1}
\ena

In the Z coordinates these solutions  will imply:
\bea
\prtb(\lambda {\cal J}^{\natural,0}_{Z,0,\star}\ZBZ) -\prt(\mu
\lambda {\cal J}^{\natural,0}_{Z,0,\star}\ZBZ)=0\nn
\prt(\lambdab \bar{\cal J}^{\natural,0}_{\bZ,0,\star}\ZBZ)-
\prtb(\mub
\lambdab \bar{\cal J}^{\natural,0}_{\bZ,0,\star}\ZBZ)=0\nn
\ena
that is:

\bea
\lambda\lambdab(1-\mu\mub)\pabZ {\cal
J}^{\natural,0}_{Z,0,\star}\ZBZ=0\nn
\lambda\lambdab(1-\mu\mub)\paZ\bar{\cal
J}^{\natural,0}_{\bZ,0,\star}\ZBZ=0\nn
\ena
which means that ${\cal J}^{\natural,0}_{Z,0,\star}\ZBZ$ is a
holomorphic
function iz Z.

This constraint can be imposed in a diff-invariant way by imposing:

\bea
\delta\biggl(\gamma^Z    {\cal J}^{0}_{Z,0,\star}\ZBZ \biggr)=0\nn
\delta\biggl(\bar{\gamma}^{\bZ}\bar{\cal J}^{0}_{\bZ,0,\star}\ZBZ
\biggr)=0
\lbl{contrs1}
\ena
or equivalently:
\bea
\delta\biggl(C {\cal{S}}^{0}_{z,0,\star}\zbz \biggr)=0
\lbl{constr21}
\ena

\bea
\delta\biggl(\bar{C}^{\bz}\bar{\cal{S}}^{0}_{\bz,0,\star}\zbz
\biggr)=0
\lbl{constr22}
\ena

Projecting \rf{constr21} into underivated ghost factors we obtain:

\bea
c \bc \biggl[ -\prtb {{\cal{S}}^{0}_{z,0,\star}}\zbz  +\prt\mu
{{\cal{S}}^{0}_
{z,0,\star}}\zbz + \mu \prt{{\cal{S}}^{0}_{z,0,\star}}\zbz \biggr]
=0 \nn
c \biggl[ \biggl(\prt c+ \mu \prt
\bc\biggr){{\cal{S}}^{0}_{z,0,\star}}\zbz
- \hat{\delta}{{\cal{S}}^{0}_{z,0,\star}}\zbz \biggr]=0 \nn
\bc \mu \biggl[ \biggl(\prt c+ \mu \prt
\bc\biggr){{\cal{S}}^{0}_{z,0,\star}}\zbz
- \hat{\delta}{{\cal{S}}^{0}_{z,0,\star}}\zbz \biggr]=0 \nn
\lbl{proconstr12}
\ena
and similarly for \rf{constr22}

We recall that the $\star$ index will indicate arbitrary "big" $ Z
$ $\bZ$
indices, so  the switching to "little" coordinates from
${{\cal{S}}^{0}_{z,0,\star}}\zbz \equiv {{\cal{S}}^{0}_{z,0,Z^n,
\bZ^m}}\zbz $
is done with a suitable $\lambda$ and
$\lambdab$ rescaling.

\bea
{\cal{S}}^{0}_{z,0,Z^n,\bZ^m}\zbz=\frac{{\cal{S}}^{0}_{z,0,z^n}\zbz}{\lambda^n
\lambdab^m}\nn
\bar{{\cal{S}}}^{0}_{\bz,0,Z^n,\bZ^m}\zbz=\frac{\bar{{\cal{S}}}^{0}_{\bz,0,
\bz^n}\zbz}{\lambda^n\lambdab^m}
\lbl{rescaling0}
\ena

It is easy to verify that the previous currents do not verify the
local current
conservation in $\mu$ and $\mub$ \rf{holocurrent} unless a
particular tensorial content
is realized: in particular this
 is achieved
in \rf{proconstr12} only if ${\cal{S}}^{0}_{z,0,\star}\zbz$ will
contain
only $Z$ indices and $\bar{{\cal{S}}}^{0}_{\bz,0,\star}\zbz$
only $\bZ$ ones, signature of the holomorphicity in Z.

So we obtain local currents ${\cal{S}}^{0}_{z,0,z^n}\zbz$
$\bar{{\cal{S}}}^{0}_
{\bz,0,\bz^n}\zbz$ defined as:

\bea
{\cal{S}}^{0}_{z,0,Z^n}\zbz=\frac{{\cal{S}}^{0}_{z,0,z^n}\zbz}{\lambda^n}\nn
\bar{{\cal{S}}}^{0}_{\bz,0,\bZ^n}\zbz=\frac{\bar{{\cal{S}}}^{0}_{\bz,0,
\bz^n}\zbz}
{\lambdab^n}
\lbl{rescaling00}
\ena

The constraint equations will be so modified:

\bea
c \bc \biggl[ -\prtb {{\cal{S}}^{0}_{z,0,z^n}}\zbz  +(n+1)\prt\mu
{{\cal{S}}^{0}_
{z,0,z^n}}\zbz + \mu \prt{{\cal{S}}^{0}_{z,0,z^n}}\zbz \biggr] =0
\nn
c \biggl[(n+1) \biggl(\prt c+ \mu \prt
\bc\biggr){{\cal{S}}^{0}_{z,0,z^n}}\zbz
- \hat{\delta}{{\cal{S}}^{0}_{z,0,z^n}}\zbz \biggr]=0 \nn
\bc \mu \biggl[(n+1) \biggl(\prt c+ \mu \prt
\bc\biggr){{\cal{S}}^{0}_{z,0,z^n}}\zbz
- \hat{\delta}{{\cal{S}}^{0}_{z,0,z^n}}\zbz \biggr]=0 \nn
\lbl{proconstr112}
\ena
and similarly for the c.c.

It is important to note that the current
${{\cal{S}}^{0}_{z,0,z^n}}\zbz$
has a definite covariance property in the sense  that is a "true"
$(n+1),0$
tensor, it has a spin value (n+1) .

In terms of holomorphic ghosts we get:

\bea
\delta{\cal{S}}^{\natural,r-1}_{z,0,z^n}\zbz=\prt\biggl(
C\zbz{\cal{S}}^{\natural,r-1}_{z,0,z^n}\zbz\biggr)+n\prt
C\zbz{\cal{S}}^{\natural,r-1}_{z,0,z^n}
\ena
\bea
\delta\bar{\cal{S}}^{\natural,r-1}_{\bz,0,\bz^n}\zbz=\prtb\biggl(\bC\zbz
\bar{\cal{S}}^{\natural,r-1}_{\bz,0,\bz^n}\zbz\biggr)+n\prtb\bC\zbz
\bar{\cal{S}}^{\natural,r-1}_{\bz,0,\bz^n}\zbz
\ena

\sect{The 1-forms in Lagrangian local Quantum Field Theory }

\subsection{ The Classical Level}

\indent

The previous Section has introduced, from a heuristic point of view
the descent
Equations which define the 1-forms. Moreover  these equations play
an important
role in the dynamics; so they have to be embedded in a Lagrangian
model whose
quantum extension will provide informations concerning the
renormalization of
those currents.

We have to introduce an invariant Classical
Action $\mbf{\Gamma}^{Cl}_0$ such that:
\bea
\delta_0\mbf{\Gamma}^{Cl}_0=0
\ena

In \cite{theone} we have shown that the more general invariant
Classical Action
under diffeomorphisms takes the form:

\bea
\mbf{\Gamma}^{Cl}_0 &=& \int \biggl(\sum_{j}
a_j [\phi_{1-j,0} (\prtb - \mu \prt - j \prt \mu)
\phi_{j,0} + c.c.]\zbz d\bz \wedge dz  \nn[2mm]
&&\hs{-10} +({1 - \mmb})(\phi_{1,0}\phi_{0,1} +c.c)\zbz d\bz \wedge
dz  \nn[2mm]
&&\hs{-10} + b_j [\phi_{-j,-j} (\prt - \mub \prtb -j \prtb \mub)
	 \phi_{0,j} \frac{\prtb - \mu \prt -j \prt \mu}{1 - \mmb}
	 \phi_{j,0} ]\zbz d\bz \wedge dz
\lbl{exaction} \\
&+& \sum_{n \geq 0} \biggl[ c_{n} [(\phi_{00})^n
(\prt - \mub \prtb) \phi_{00} \frac{\prtb - \mu \prt}
{1 - \mmb} \phi_{00} ]\zbz +c.c. \biggr]d\bz \wedge dz \biggr)
\nonumber
\ena

We shall treat here, for the sake of simplicity, the spin zero
case;  in
this case the most general invariant Lagrangian  reads:
\bea
\mbf{\Gamma}^{Cl}_0 &=& \int \biggl(
 \sum_{n \geq 0} \biggl[ c_{n} [(\phi_{00})^n
(\prt - \mub \prtb) \phi_{00} \frac{\prtb - \mu \prt}
{1 - \mmb} \phi_{00} ]\zbz d\bz \wedge dz \biggr] \biggr) \nonumber
\lbl{action11}
\ena

We remark that, due to the dimensionless character of the scalar
field,
the only diff-invariance requirement will imply an infinite number
of
interaction terms at the Classical level, raising a lot of problems
on the
physical meaning of the model; anyhow several criteria can be
established on
the resummation of the interacting part, which will involve
particular
addition conditions on the definition on the model which would not
destroy
the reparametrization invariance. These aspects do not compromise
our
treatment which will hold validity for all these classes of models.

According to the general prescription, we have to introduce the
B.R.S. variations coupled to $\Phi,\Pi$ negative charged external
fields.

Furthermore, as said before,  we put into the dynamics the 1-forms
current
coupled to external fields.

\bea
\mbf{\Gamma}^{Cl} &=&\mbf{\Gamma}^{Cl}_0+\mbf{\Gamma} ^{Cl}_{1}+
\mbf{\Gamma}^{Cl}_{2}\nn
\ena
where the "source" Action reads:
\bea
\mbf{\Gamma}^{Cl}_{1} &=&\int d\bz \wedge dz\biggl(
\biggl(J_{1,1}\zbz \phi_{0,0}\zbz + \gamma_{1,1}\zbz s
\phi_{0,0}\zbz\biggr) \nn
&+& \eta\zbz s \mu\zbz + \bar{\eta}\zbz s \bar{\mu}\zbz
+ \zeta\zbz s c\zbz + \bar{\zeta}\zbz s \bc\zbz \biggr)\biggr)
\lbl{action111}
\ena
whith:

\eq
\begin{array}{rl}
s \mu &=\ \cdotd \mu - \mu (\prt c + \mu\prt\bc) + \prtb c +
\mu\prtb\bc \\[2mm]
sc &=\ \cdotd c \\[2mm]
s \phi_{0,0} &=\ \cdotd \phi_{0,0} \ ,
\end{array}
\lbl{sLOC}
\en
and of course the complex conjugate expressions and the "current"
Action :
\bea
\mbf{\Gamma}^{Cl}_{2}
&=&\int d\bz \wedge dz\biggl(
\rho^{z^n}\zbz \Lambda^{-1}_{z,\bz, z^n}\zbz
+\bar{\rho}^{\bz^n}\zbz\bar{\Lambda}^{-1}_{z,\bz, \bz^n}\zbz \nn
&+&\beta_{\bz}^{z^n} \zbz {{\cal{S}}^{0}_{z,0,z^n}\zbz}
+{\bar{\beta}}_{z}^{\bz^n}\zbz
{\bar{\cal{S}}}^{0}_{\bz,0,\bz^n}\zbz\biggr)
\lbl{actionbeta111}
\ena

We have to impose the invariant condition:
\bea
\delta_0\mbf{\Gamma}^{Cl}=0
\ena

The U.V. dimensions of the constituents of the model are:

\bea
[\prt]=[\prtb]=1 \nn
\\[1mm]\nonumber
[\phi_{0,0}]=0 \nn
\\[1mm]\nonumber
[\mu]=[c]=[\bc]=0\nn
\\[1mm]\nonumber
[\gamma_{1,1}]=1\nn
\\[1mm]\nonumber
[\zeta]=[\eta]
=[\bar{\eta}]=1\nn
\\[1mm]\nonumber
[\rho^{z^n}]=
[\beta_{\bz}^{z^n}]=[\bar{\beta}_{z}^{\bz^n}]=1-n\nn
\\[1mm]\nonumber
[{{\cal{S}}^{0}_{z,0,z^n}\zbz}]=1+n\nn
\\[1mm]\nonumber
\lbl{dim}
\ena

The external fields coupled to the holomorphic currents are
introduced in the
Lagrangian, by fixing their variations in order to get the descent
equations
seen in the previous Section:
So we have:

\bea
s \rho^{z^n} \zbz &=\ \cdotd \rho^{z^n}\zbz-n(\prt c\zbz
+\mu\prt\bc\zbz)
\rho^{z^n}\zbz  \nn
\lbl{srho}
\ena
\bea
s \beta_z^{z^n}\zbz & = & \cdotd \beta_z^{z^n}\zbz\nn
& & +(\prt c\zbz  +\prtb\bc\zbz)
\beta_z^{z^n}\zbz\nn
& &- (n+1) (\prt c\zbz+\mu\prt\bc)
\beta_z^{z^n}\zbz\nn
& & +\prtb \rho^{z^n} \zbz
- \mu\prt\rho^{z^n} \zbz
+n\prt\mu\rho^{z^n} \zbz\nn
\lbl{sbeta}
\ena
and their c.c. expressions. That is, if we define:
\bea
\rho^{Z^n} \ZBZ={(\lambda)}^n \rho^{z^n} \zbz
\ena

\bea
\beta_Z^{Z^n}\ZBZ=\frac{\lambda^{(n+1)}}{\lambda\lambdab(1-\mu\mub)}
\beta_z^{z^n}\zbz
\ena
we get the following descent equations for the sources:

\bea
\hat{\delta}\beta_Z^{Z^n}\ZBZ=\pabZ\rho^{Z^n} \ZBZ\nn
\hat{\delta}\rho^{Z^n} \ZBZ=0
\ena
the introduction of the  previous fields allows us to reproduce at
the Lagrangian
level the right properties of the holomorphic currents
${{\cal{S}}^{0}_{\bz,0,z^n}\zbz}$ at the classical level, when the
simmetry is preserved.

The role of the $\rho^{z^n}$ field, as inhomogeneous part
of the $\beta$ tranformations,
is of prime importance: it will
fix the current conservation
\rf{holocurrent}.
The BRS philosophy forces us to fix their
covariance properties, so the $\Lambda_{z,\bz, z^n}\zbz$ term  is a
priori
needed at the tree level: we shall show that this term is
unessential at
the Classical level, but, on the other hand, is fundamental at the
quantum
level.

The BRS operator is defined:

\bea
\delta_0 &=& \int d\bz \wedge dz \left(
\frac{\delta\mbf{\Gamma}^{Cl}}{\delta \gamma_{1,1}\zbz}
\frac{\delta}{\delta \phi_{0,0}\zbz}
+ \frac{\delta\mbf{\Gamma}^{Cl}}{\delta \phi_{0,0}\zbz}
\frac{\delta}{\delta \gamma_{1,1}\zbz} \right. \nn
&+& \frac{\delta\mbf{\Gamma}^{Cl}}{\delta \eta\zbz}
\frac{\delta}{\delta \mu\zbz}
+ \frac{\delta\mbf{\Gamma}^{Cl}}
{\delta \mu\zbz}\frac{\delta}{\delta \eta\zbz}
+ \frac{\delta\mbf{\Gamma}^{Cl}}{\delta \bar{\eta}\zbz}
\frac{\delta}{\delta \bar{\mu}\zbz} \nn
&+& \frac{\delta\mbf{\Gamma}^{Cl}}{\delta \bar{\mu}\zbz}
\frac{\delta}{\delta \bar{\eta}\zbz}
+ \frac{\delta\mbf{\Gamma}^{Cl}}{\delta \zeta\zbz}
\frac{\delta}{\delta  c\zbz}
+ \frac{\delta\mbf{\Gamma}^{Cl}}{\delta  c\zbz}
\frac{\delta}{\delta \zeta\zbz} \nn
&+&  \frac{\delta\mbf{\Gamma}^{Cl}}{\delta \bar{\zeta}\zbz}
\frac{\delta}{\delta \bc\zbz}
+ \frac{\delta\mbf{\Gamma}^{Cl}}{\delta \bc\zbz}
\frac{\delta}{\delta \bar{\zeta}\zbz}\nn
&+& s \rho^{z^n}\zbz\frac{\delta}{\delta {\rho}^{z^n}\zbz}
+ s\bar{\rho}^{\bz^n}\zbz\frac{\delta}{\delta
\bar{\rho}^{\bz^n}\zbz}
+ s\beta_{z}^{z^n}\zbz\frac{\delta}{\delta {\beta_z^{z^n}}\zbz}\nn
&+&\left.s\bar{\beta}_{z}^{\bz^n}\zbz)\frac{\delta}{\delta
{\bar{\beta}_{z}
^{\bz^n}}\zbz}
\right)
\lbl{brsbeta}
\ena
with
\bea
s \gamma_{1,1}\zbz &=&\frac{\delta\mbf{\Gamma}^{Cl}_0}{\delta
\phi_{0,0}\zbz}
+\cdotd\gamma_{1,1}\zbz + (\prt c\zbz + \prtb \bc\zbz)
\gamma_{1,1}\zbz
 \nn[2mm]
\lbl{BRScurrent1}
\ena
and
\bea
&&\hs{-30} s \eta\zbz =
\frac{\delta\mbf{\Gamma}^{Cl}_0}
{\delta \mu\zbz}+
\cdotd\eta\zbz + (\prt c\zbz + \prtb \bc\zbz) \eta\zbz\nn
&&\hs{-30}+ (\prt c\zbz +2 \mu\zbz\prt\bc\zbz-
\prtb\bc\zbz)
\eta\zbz \nn[2mm]
\lbl{BRScurrent2}
\ena

\bea
&&\hs{-30} s \zeta\zbz =-\gamma_{1,1}\zbz\prt\phi_{0,0}\zbz
-\eta\zbz\prt\mu\zbz-\prt(\mu\zbz\eta\zbz)-\prtb\eta\zbz\nn
&&\hs{-30}-\bar{\eta}\zbz\prt\mub\zbz-\prt(\mu^2\zbz\bar{\eta}\zbz)
+\zeta\zbz\prt c\zbz-\prt(\zeta\zbz c\zbz)-\prtb(\zeta \bc\zbz)\nn
\lbl{BRScurrent3}
\ena

So we can write the current Ward identities, coming from the B.R.S.

variation of the external  sources
$\rho^{z^n}\zbz$,$\beta_{\bz}^{z^n} \zbz$.
They reproduce in a functional approach the descent equations
just encountered in the previous Section, written in terms of
$\Lambda_{z,\bz, z^n}\zbz$,${{\cal{S}}^{r-1}_{z,0,z^n}\zbz}$ and
their c.c.

\bea
\frac{\delta}{\delta \rho^{z^n}
\zbz}{\biggl(\delta_0\mbf{\Gamma}^{Cl}\biggr)}
=\delta\frac{\delta_0\mbf{\Gamma}^{Cl}}{\delta \rho^{z^n} \zbz}
-\prt\biggl(c\frac{\delta\mbf{\Gamma}^{Cl}}{\delta \rho^{z^n}
\zbz}\biggr)
-\prtb\biggl(\bc\frac{\delta\mbf{\Gamma}^{Cl}}{\delta \rho^{z^n}
\zbz}\biggr)\nn
-n(\prt c\zbz +\mu\prt\bc\zbz)\frac{\delta\mbf{\Gamma}^{Cl}}{\delta
\rho^{z^n
} \zbz}\nn
-\prtb \frac{\delta\mbf{\Gamma}^{Cl}}{\delta{\beta}_{\bz}\zbz}
+\mu\prt\frac{\delta\mbf{\Gamma}^{Cl}}{\delta{\beta}_{\bz}\zbz}
+(n+1)\prt\mu\frac{\delta\mbf{\Gamma}^{Cl}}{\delta{\beta}_{\bz}\zbz}\nn
\equiv
\biggl({\delta_0}- c \prt-\bc \prtb-\prtb  \bc-\prt c \biggr)
\Lambda_{z,\bz, z^n}\zbz\nn
-\prtb {{\cal{S}}^{0}_{z,0,z^n}}\zbz  +(n+1)\prt\mu {{\cal{S}}^{0}_
{z,0,z^n}}\zbz + \mu \prt{{\cal{S}}^{0}_{z,0,z^n}}\zbz=0
\lbl{ward10}
\ena

\bea
\frac{\delta}
{\delta \beta_{\bz}^{z^n}
\zbz}\biggl(\delta_0\mbf{\Gamma}^{Cl}\biggr)
=\delta_0
\frac{\delta\mbf{\Gamma}^{Cl}}{\delta \beta_{\bz}^{z^n} \zbz}
-\prt\biggl( c\zbz \frac{\delta\mbf{\Gamma}^{Cl}}{\delta
\beta_{\bz}^{z^n} \zbz}\biggr)
-\prtb\biggl( \bc\zbz \frac{\delta\mbf{\Gamma}^{Cl}}{\delta
\beta_{\bz}^{z^n}
\zbz}
\biggr)\nn
+(\prt c\zbz +\prtb\bc\zbz)\frac{\delta\mbf{\Gamma}^{Cl}}{\delta
\beta_
{\bz}^{z^n \zbz}} -(n+1)(\prt c+\mu\prt\bc)
\frac{\delta\mbf{\Gamma}^{Cl}}{\delta {\beta}_{\bz}^{z^n \zbz}}\nn
\equiv\biggl(\widehat{\delta}- c \prt-\bc \prtb-(n+1)(\prt
c+\mu\prt\bc)\biggr
)
{{\cal{S}}^{r-1}_{z,0,z^n}\zbz}=0 \nn
\lbl{ward11}
\ena

Their solutions are carried out as before; introducing:

\bea
{{\cal{S}}^{0}_{Z,0,Z^n}\ZBZ}=\frac{{{\cal{S}}^{0}_{z,0,z^n}\zbz}}{\lambda
^{(n+1)}}\nn
\Lambda_{Z,\bZ, Z^n}\ZBZ=\frac{\Lambda_{z,\bz,
z^n}\zbz}{\lambda^{(n+1)}\lambdab
(1-\mu\mub)}\nn
\lbl{redefinition1}
\ena
they become:
\bea
\hat{\delta}{{\cal{S}}^{0}_{Z,0,Z^n}\ZBZ}=0\nn
\hat{\delta}\Lambda_{Z,\bZ,
Z^n}\ZBZ=\pabZ{{\cal{S}}^{0}_{Z,0,Z^n}\ZBZ}\nn
\lbl{descent22}
\ena
so:

\bea
{{\cal{S}}^{0}_{Z,0,Z^n}\ZBZ}={{\cal{S}}^{0,\natural}_{Z,0,Z^n}\ZBZ}+
\hat{\delta}{\widehat{\cal{S}}^{r-2}_{Z,0,Z^n}\ZBZ}
\lbl{descent221}
\ena
so \rf{ward10} is rewritten as
\bea
\hat{\delta}\biggl(\Lambda_{Z,\bZ,
Z^n}\ZBZ-\paZ{\widehat{\cal{S}}^{r-2}_
{Z,0,Z^n}\ZBZ}\biggr)=\pabZ{{\cal{S}}^{0,\natural}_{Z,0,Z^n}\ZBZ}
\lbl{descent222}
\ena

But,since, $\pabZ{{\cal{S}}^{0,\natural}_{Z,0,Z^n}\ZBZ}$ is an
element of the $\hat{\delta}$-cohomology, the previous equation is
consistent only if each term is identically zero: we have shown so
that the
diff invariance will imply the holomorphicity of
${{\cal{S}}^{0,\natural}_{Z,0,Z^n}\ZBZ}$, that is:

\bea
\frac{\delta}{\delta \rho^{z^n}
\zbz}{\biggl(\delta_0\mbf{\Gamma}^{Cl}\biggr)}=0
\lra
\pabZ{{\cal{S}}^{0,\natural}_{Z,0,Z^n}\ZBZ}=0
\lbl{descent223}
\ena

So the current conservation will derive from the inhomogeneous part
of the
$\beta$ variation, so a priori we have to require:
\bea
 (\prtb \rho^{z^n} \zbz
- \mu\prt\rho^{z^n} \zbz +n\prt\mu\rho^{z^n} \zbz) \neq 0 .
\lbl{rhocondition}
\ena
the eq.\rf{rhocondition} will imply:

\bea
\pabZ\rho^{Z^n} \zbz\neq 0
\lbl{rhocondition1}
\ena

On the other hand it is easy to realize from \rf{ward10}
\rf{ward11}
that in the quantum extension of the
model, the possible $\rho$ and $\beta$ dependent anomalies will
spoil the current
conservation and their covariance properties respectively.

The B.R.S. approach consists in the study of the cohomology of the
$\delta$
operator in the space of local functionals, the charge zero space
will identify
the Classical Action while the charge one will give the quantum
anomalies.

This analysis has to be done as the one carried out in
\cite{theone},
where, in a similar way  we have related the diff-mod {\bf{d}}
cohomolology space to the one  of $\delta$ within the class of
local functions.

Calling  $\DDEL{p}{2}$ the more general element of the diff-mod
{\bf{d}}
cohomolology and labeling with the $\natural$ index the $\delta$
cohomology
elements we can find the 2-form extension of \rf{coho} as
calculated in
\cite{theone}

\bea \leqa{
\DDEL{p}{2} = \delnat{p}{2} - \DER{c} \DER{\bc} \delnat{p+2}{0}
dz \wedge d\bz } \nn
&&\mbox{}\lbl{coho2form}\\[-2mm]
&& - \DER{c} \delnat{p+1}{1} dz - \DER{\bc} \delnat{p+1}{1} d\bz
- d \delhat{p}{1} + \delta \delhat{p-1}{2}\ ,\nonumber
\ena

The novelty of this paper with respect to \cite{theone} consists in

introducing  the currents inside the dynamics of the Lagrangian,
and
(as it is easy to realize from \rf{ward10} \rf{ward11} )
the quantum extension of the
model might generate $\rho$ and $\beta$ dependent anomalies
which could spoil the current
conservation and their covariance properties respectively, as
already stated.

The next Section will investigate this possibility.

\subsection{ The Quantum Level}

\indent

The quantum extension of the model has to be done as in our paper
\cite{theone};
first of all  we have to parametrize the anomaly as:

\bea
\delta\mbf{\Gamma}&=&\Delta\nn
&=&\int d\bz \wedge dz
\Delta^0 \zbz  +\sum_n \biggl(\rho^{z^n}\zbz \Delta^{0,n}_{z, \bz,
z^n} \zbz\nn
&+&\bar{\rho}^{\bz^n}\zbz \bar{\Delta}^{0,n}_{z, \bz, \bz^n} \zbz
+\bar{\beta}_z^{\bz^n} \zbz\bar{\Delta}^{1,n}_{\bz\bz^n}\zbz\nn
&+&{\beta}_{\bz}^{z^n} \zbz\Delta^1_{z,z^n}\zbz\biggr)
\lbl{qap}
\ena
and $\Delta^{0,n}_{z, \bz, z^n} \zbz$ has $\Phi, \Pi$ charge equal
to zero, and
$\Delta^{1,n}_{z,z^n}$ with charge one, they both have U.V.
dimensions 2+n.

The hunting of anomalies has to be done as in \cite{theone}, and
within this model
we show in the Appendix that:

{\bf Theorem}
{\em
 The ghost sectors ($\Phi$-$\Pi$ charge sectors) of the
$\tld{\delta}$-cohomology in the space of analytic functions of the
fields,
where the fields $\lambda$ and $\lambdab$  satisfy
the equation \rf{intfac}, and completed with the terms
$\{\ln\lambda,\ln\lambdab\}$,seen as independent fields
depend only on terms containing underivated source
$\rho$ which multiply zero ghost sector elements of the cohomology.

The zero ghost sector is, on the other hand, non-trivial only in
the part
which contains matter fields. Its elements will contain
no free $z$ and $\bz$ indices, i.e. they are "scalar-like"
quantities with
respect to "little indices" but can hold the tensorial content with
respect
the "big indices" $Z$ and $\bZ$.
A generic element of
this space will be a $(h,\bar{h})$-conformal quantity of the form
\[ [f( \prt^m_Z \prt^n_{\bZ} \phi_{0,0} \ZBZ )]_{h,\bar{h}},\]
where  $f$ is an analytic function, (polynomial).
}

\bigskip

So we are left with a $\rho$ dependent anomaly:

\bea
\Delta^{\natural}_{z\bz}\zbz=\lambda\lambdab(1-\mu\mub)\rho
\Delta^{2\natural}_{Z\bZ}\zbz
\lbl{decompanom6}
\ena
but from the transformation laws \rf{srho} we argue that:

\eq
\begin{array}{rl}
\rho\zbz \approx \prt c \zbz +\prtb \bc \zbz\nn
\end{array}
\lbl{rho}
\en
so eq \rf{decompanom6} recover an ordinary trace anomaly which can
be
reabsorbed by a counterterm

\ceq
\int d\bz \wedge dz \rho\zbz\phi_{0,0}\zbz \gamma_{1,1}\zbz
\cen

Indeed the B.R.S. variation of the previous term gives the anomaly:
\bea \leqa{
s \int d\bz \wedge dz \rho\zbz
\phi_{0,0}\zbz \gamma_{1,1}\zbz =} \nn
& &- \int d\bz \wedge dz \rho\zbz \phi_{0,0}\zbz
\frac{\delta \mbf{\Gamma}^{Cl}_0}{\delta\phi_{0,0}\zbz}
\lbl{canc1}
\ena

At this stage we have to reconsider the model and all the
cohomolology
calculations serve  to recover every possible origins of anomalies:
in Ref
\cite{theone} the locality requirement, which forces the
elimination of
our vectorial space $\cal{V}$  of  $\lambda$, $\lambdab$ (but
more important of $\ln \lambda$ and $\ln \lambdab$) dependence, was
the origin
of the holomorphic anomaly, which is represented by  $\Delta^0 \zbz
$.
In this
paper we are involved in the ${\beta}_{z}^{z^n} \zbz$ and
$\rho^{z^n}\zbz$
andtheir c.c. external dependent anomalies, so
we have to analyze again their disappearance from the cohomology
sectors.

We recall that the cancellation of the $\rho^{z^n}\zbz$ anomalies
(see Appendix)
are governed by the term
\bea
 (\prtb \rho^{z^n} \zbz
- \mu\prt\rho^{z^n} \zbz +n\prt\mu\rho^{z^n} \zbz) \neq 0 .
\lbl{rhocondition11}
\ena
so possible $\rho^{z^n} \zbz$ dependent anomalies can appear in the
slice where
\bea
 (\prtb \rho^{z^n} \zbz
- \mu\prt\rho^{z^n} \zbz +n\prt\mu\rho^{z^n} \zbz) = 0 .
\lbl{rhocondition12}
\ena
that is:
\bea
\pabZ\rho^{Z^n} \ZBZ= 0
\lbl{rhocondition111}
\ena

If \rf{rhocondition12} is valid, the field $\beta_z^{z^n}\zbz$
is a "true" tensor density
\bea
s \beta_z^{z^n}\zbz & = & \cdotd \beta_z^{z^n}\zbz\nn
& & +(\prt c\zbz  +\prtb\bc\zbz)
\beta_z^{z^n}\zbz\nn
& &- (n+1) (\prt c\zbz+\mu\prt\bc)
\beta_z^{z^n}\zbz\nn
\lbl{sbeta1}
\ena

Furthermore in this region the covariance properties of
${{\cal{S}}^{0,
\natural}_{Z,0,Z^n}\zbz}$ are not modified, due \rf{ward11} (or if
you
prefer the second and third \rf{proconstr112} condition),
and no $\beta_z^{z^n}\zbz$ dependent anomalies  can appear.

With this constraint the BRS variation of
$\rho^{z^n} \zbz$  becomes
\bea
s\rho^{z^n} \zbz= C\zbz\prt\rho^{z^n} \zbz-n\prt C\zbz\rho^{z^n}
\zbz
\lbl{rhocondition1111}
\ena
It is so evident that this is a sign of the holomorphic
factorization,
so the anomalies have to be searched as elements of the local
cohomology
of the $\delta$ operator as polynomial in $\rho^{z^n} \zbz$ and its

$\prt$ derivatives, with the remaining field content in order to
get
$\Phi.\Pi$ charge and U.V.dimension equal to 3.

\bea
\Delta^{\natural}\zbz=\prt_r\rho^{z^n} \zbz\Delta^{r}_n\zbz \nn
(\prt_r\equiv \prt_1\prt_2\cdots\prt_r)\nn
\lbl{rhocondition1112}
\ena

\bea
dim\Delta^{r}_n\zbz =2+n-r
\lbl{rhocondition1113}
\ena
so for  $r\leq 2+n$ the cocycle condition:
\bea
s\Delta^{\natural}\zbz=0
\ena
implies:
\bea
\prt_r\biggl(C\zbz\prt\rho^{z^n}
\zbz-n\prt C\zbz\rho^{z^n} \zbz\biggr)-\prt_r\rho^{z^n} \zbz
s\Delta^{r}_n\zbz=0
\lbl{rhocondition1114}
\ena
and then:
\bea
\sum_{j\leq r} \prt_s\rho^{z^n} \zbz\biggl[\left(\begin{array}{c}
r\\j
\end{array}\right)\biggl(-\prt_j C \zbz \delta^{s}_{r-j+1} + n
\prt_{j+1} C \zbz
\delta^{s}_{r-j}\biggr)\Delta^{r}_n\zbz-\delta^{s}_{r}
s\Delta^{r}_n\zbz
\biggr]=0
\lbl{rhocondition1115}
\ena

In other words:
\bea
\left(\begin{array}{c} r\\j
\end{array}\right)\biggl(-\prt_j C \zbz \delta^{s}_{r-j+1} + n
\prt_{j+1} C \zbz
\delta^{s}_{r-j}\biggr)\Delta^{r}_n\zbz=\delta^{s}_{r}
s\Delta^{r}_n\zbz
\lbl{rhocondition1116}
\ena
for each $s,n\geq 0$

For s=r

\bea
s\Delta^{r}_n\zbz=(n-r)\prt C\zbz \Delta^{r}_n\zbz
\lbl{rhocondition1117}
\ena
and for $s\neq r$

\bea
\sum_{j\leq r,j\neq 1} \left(\begin{array}{c} r\\j
\end{array}\right)\biggl(-\prt_j C \zbz \delta^{s}_{r-j+1} + n
\prt_{j+1} C \zbz \delta^{s}_{r-j}\biggr)\Delta^{r}_n\zbz=0
\lbl{rhocondition1118}
\ena

Now the BRS operator will always  contain $c\prt+\bc\prtb\cdots$
which do not appear in \rf{rhocondition1117}
so the only solution will be for
\bea
n=r
\ena
 and the anomaly takes the form
\bea
\Delta^{\natural}\zbz=\prt_n\rho^{z^n} \zbz\Delta^{n}\zbz
\lbl{rhocondition11121}
\ena
where:

\bea
dim\Delta^{n}\zbz =2
\lbl{rhocondition11131}
\ena
for each n, and:
\bea
s\Delta^{n}\zbz=0
\lbl{rhocondition11171}
\ena

Furthermore from \rf{rhocondition1118}, for each $n$, we have:
\rf{rhocondition1118}
\bea
 - C \zbz  \left(\begin{array}{c} n-1\\0
\end{array}\right)\Delta^{n-1}\zbz \nn
+\prt\prt C\zbz\biggl(-\left(\begin{array}{c} n+1\\2
\end{array}\right)+(n+1)\left(\begin{array}{c} n+1\\1
\end{array}\right)\biggr)\Delta^{n+1}\zbz\nn
+\prt\prt\prt C\zbz\biggl(-\left(\begin{array}{c} n+2\\3
\end{array}\right)+(n+2)\left(\begin{array}{c} n+2\\2
\end{array}\right)\biggr)\Delta^{n+2}\zbz\nn
+\cdots=0
\lbl{rhocondition111811}
\ena
where dots will contain $\Delta^{n}\zbz$ $(n\geq 3)$ with fixed
dimensions equal to 2
which multiply terms $\prt^l C\zbz$ ,$ (l\geq 4)$ .

So the only cancellation mechanism relies on $\Phi\Pi$ tricks , and

by considering that $\Delta^{n}\zbz$ does not contain any external
field,
power counting arguments forbid any solution; so:

\bea
\Delta^{n}\zbz=0\quad
for\quad
n\geq 2
\ena

Elementary considerations show that the previous conditions are all
verified only for

\bea
\Delta^{n}\zbz=C\zbz\prt\prt C\zbz,\quad
n= 0,1
\ena
so we have

\bea
\Delta^{\natural}_0\zbz=\rho\zbz C\zbz\prt\prt C\zbz=\rho\zbz s
\biggl(\prt
C\zbz\biggr)\nn
\Delta^{\natural}_1\zbz=\prt\rho^{z}\zbz C\zbz\prt\prt
C\zbz=\prt\rho^{z}\zbz s
\biggl(\prt C\zbz\biggr)\nn
\ena
since, in the case of \rf{rhocondition111},
$\rho\approx \prt C\zbz$ it is easy to realize that
$\Delta^{\natural}_0\zbz$ is a mimic of the Feigin Fuks cocycle
\cite{GF}.

In this framework we can verify that:

\bea
\Delta^{\natural}_0\zbz=s \biggl(\rho\zbz C\zbz\prt \ln
\lambda\zbz\biggr)\nn
\Delta^{\natural}_1\zbz=s \biggl(\prt\rho^{z}\zbz C\zbz\prt \ln
\lambda\zbz
-C\zbz\prt \ln \lambda\zbz \rho^{z}\zbz\prt\ln
\lambda\zbz\biggr)\nn
\ena
so they are coboundaries in "non local " basis, while in a local
ones the
compensation mechanism is not possible and give rise to anomalies.

Finally  calculations are concluded by deriving the Ward anomalies:

the Ward identity obstruction to the (1,0) current conservation
takes
the form:
\bea
\frac{\delta}{\delta \rho\zbz} \int \frac{\partial}{\partial
c(z',\bz')}
\frac{\partial}{\partial \bc(z',\bz')}\Delta^{\natural}_0
(z',\bz')  dz' \wedge d\bz'\approx\prt^2 \mu\zbz
\ena
and the (2,0) obstruction which corresponds to that of the energy
momentum tensor reads:

\bea
\frac{\delta}{\delta \rho^z\zbz} \int \frac{\partial}{\partial
c(z',\bz,)}
\frac{\partial}{\partial \bc(z',\bz')}\Delta^{\natural}_1
(z',\bz')  dz' \wedge d\bz'\approx\prt^3 \mu\zbz
\ena

\sect{Conclusions}

\indent

In the present work we have studied the role of the diffeomorphism
current
both at the Classical and at the Quantum level by computing some
specific
cohomologies.

It  has been shown, within the Beltrami parametrization of compex
structures,
that the holomorphic properties play a fundamental role in the
dynamics of
simple conformal models.
This fact again infers the relevance of the complex structure of
the Riemann
surface on which the field theoretical model is built on.

The locality requirements govern deeply the occurrence of anomalies
at the
quantum level.

The study of diffeomorphism current is not completed here and
deserves some
more careful results in particular in the meaning of the anomalous
Ward
identities for correlation functions with diffeomorphism current
insertions.

\newpage

\appendix
\renewcommand{\thesection}{Appendix \Alph{section}}
\renewcommand{\theequation}{\Alph{section}.\arabic{equation}}

\sect{The  $\tld{\delta}$  cohomology}

\indent

The previous results heavy rely on the calculation of
the $\tld{\delta}$  operator on the
space $\cal{V}$ of the local functions with positive power on the
matter field
$\phi_{0,0}$ and the $\Phi.\Pi.$ charged fields, and analitical in
the Beltrami fields.
These constaints are required on the basis of $\Phi.\Pi.$ charge
superselection rules and Lagrangian contruction and play a relevant
role, since
the cohomology definition depends not only on the operator but on
its domain
too.For the construction given in the text the space $\cal{V}$ do
not contain
underivated $c\zbz$, $\bc\zbz$ $\Phi.\Pi.$ ghosts.

The operator $\tld{\delta}$ is defined from ${\delta}$ as:

\eq
\tld{\delta} = \delta - (c \zbz \! \cdot \! \prt) =
\delta - c \zbz \grt{\{} \delta , \DER{c} \grt{\}}
- \bc \zbz \grt{\{} \delta , \DER{\bc} \grt{\}}\ ,
\lbl{a1}
\en
where:

\bea
\delta &=& sZ \zbz \DER{Z} + s\bZ \zbz \DER{\bZ} \nn[2mm]
&&+ \sum_{m,n \geq 0} \left( \prt^m \prtb^n s \phi_{j,\jb} \zbz
\DER{\prt^m \prtb^n \phi_{j,\jb}}
+ \prt^m \prtb^n s \mu\zbz \DER{\prt^m \prtb^n \mu}\right.
\nn[2mm]
&&+ \prt^m \prtb^n s \mub \zbz \DER{\prt^m \prtb^n \mub}
+ \prt^m \prtb^n s \lambda \zbz \DER{\prt^m \prtb^n
\lambda}\nn[2mm]
&&\mbox{}\lbl{brs}\\[-2mm]
&&+ \prt^m \prtb^n s \lambdab \zbz \DER{\prt^m \prtb^n \lambdab}
+ \prt^m \prtb^n s \gamma \zbz \DER{\prt^m \prtb^n \gamma} \nn[2mm]
&&+ \prt^m \prtb^n s \eta \zbz \DER{\prt^m \prtb^n \eta}
+ \prt^m \prtb^n s \bar{\eta} \zbz \DER{\prt^m \prtb^n
\bar{\eta}}\nn[2mm]
&&+ \prt^m \prtb^n s \zeta \zbz \DER{\prt^m \prtb^n \zeta}
+ \prt^m \prtb^n s \bar{\zeta} \zbz \DER{\prt^m \prtb^n
\bar{\zeta}}\nn[2mm]
&&+ \prt^m \prtb^n s \rho \zbz \DER{\prt^m \prtb^n \rho}
+ \prt^m \prtb^n s \beta \zbz \DER{\prt^m \prtb^n \beta}\nn[2mm]
&&+ \prt^m \prtb^n s \bar{\beta} \zbz \DER{\prt^m \prtb^n
\bar{\beta}}
+ \prt^m \prtb^n s \bar{\rho} \zbz \DER{\prt^m \prtb^n
\bar{\rho}}\nn[2mm]
&& \left. + \prt^m \prtb^n sc \zbz \DER{\prt^m \prtb^n c}
+ \prt^m \prtb^n s\bc \zbz \DER{\prt^m \prtb^n \bc} \right) \ .
\nonumber
\lbl{delta00}
\ena

We recall that $\tld{\delta}$ acts on the space on the fields and
their
derivatives considered as independent coordinates, as they stand
for local
Fock representation of the model.

The Spectral Sequencie analysys \cite{leray,dixon,beppe} is a
"perturbative-like" method method which
allows to recover, by recursion, a space which is isomorphic to the
cohomology
one.

First of all an adjoint procedure is introduced into the game (just
copying
the Fock-like creation and destruction procedure \cite{dixon})
by the formal replacement \cite{beppe}
of the formal derivative with respect to the field and their
derivatives by
the formal multiplication with respect the same quantities and vice
versa.

Introducing the self-adjoint operator:

\eq
\nu = \sum_{m,n\geq 0,\ m+n \geq 1} (m+n) \left(
\prt^m \prtb^n c \zbz \DER{\prt^m \prtb^n c} +
\prt^m \prtb^n \bc \zbz \DER{\prt^m \prtb^n \bc} \right)\ ,
\lbl{nu}
\en
whose eigenvalues provide the counting of the order of the ghost
derivatives;
so the space $\cal{V}$ can be decomposed into a direct sum of
subspaces;
futhermore $\tld{\delta}$ can be graded with respect to $\nu$ as:

\eq
\grt{[} \nu , \tld{\delta} \grt{]} = \sum_{m,n\geq 0,\ m+n\geq 1}
(m+n)\,
\tld{\delta}(m+n)\ ,
\lbl{filtra}
\en

In general the Spectral Sequence method insures that the
$\tld{\delta}$
cohomology is isomorphic to the solutions $\tld{\Delta} \zbz$ of
the system:

\eq \left\{ \begin{array}{c}
\tld{\delta}(m+n) \tld{\Delta} \zbz=0 \\[2mm]
\tld{\delta}^{\dag}(m+n) \tld{\Delta} \zbz=0
\end{array} \right. \ .
\lbl{iso}
\en
or (in other words)  $\tld{\Delta} \zbz$ are zero modes of the
Laplacians
$\left\{\tld{\delta}(m+n),\tld{\delta}^{\dag}(m+n)\right\}$, such
that:

\bea
\left\{\tld{\delta}(m+n),\tld{\delta}^{\dag}(m+n)\right\}
\tld{\Delta} \zbz=0
\ena

The first level of filtration  will select the part of the operator
which
does not contain any $\Phi,\Pi$ fields:

\bea
\tld{\delta}(0)=\sum_{m,n \geq 0} \left\{  \left(
\prt^m \prtb^n \frac{\delta\mbf{\Gamma}^{Cl}_0}{\delta
\phi_{0,0}\zbz} \zbz \DER{\prt^m \prtb^n
\gamma_{1,1}}\right.\right.
 \nn[2mm]
+ \prt^m \prtb^n \frac{\delta\mbf{\Gamma}^{Cl}_0}{\delta
\mu\zbz} \DER{\prt^m \prtb^n \eta}
+\prt^m \prtb^n \frac{\delta\mbf{\Gamma}^{Cl}_0}{\delta
\mub\zbz} \zbz \DER{\prt^m \prtb^n \bar{\eta}} \\[2mm]\nonumber
+ \prt^m \prtb^n \biggl((-\gamma_{1,1}\zbz\prt
\phi_{0,0}\zbz
-\eta\zbz\prt\mu\zbz-\prt(\mu\zbz\eta\zbz)-\prtb\eta\zbz\\[2mm]\nonumber
-\bar{\eta}\zbz\prt\mub\zbz-\prt(\mu^2\zbz\bar{\eta}\zbz)
 \biggr) \DER{\prt^m \prtb^n \zeta}  \\[2mm]\nonumber
+ \prt^m \prtb^n \biggl((-\gamma_{1,1}\zbz
\prtb\phi_{0,0}\zbz
-\bar{\eta}\zbz\prtb\mub\zbz-\prtb(\mub\zbz\bar{\eta}\zbz)-\prt\bar{\eta}
\zbz\\[2mm]\nonumber
-{\eta}\zbz\prtb\mu\zbz-\prtb(\mub^2\zbz{\eta}\zbz)
 \biggr) \DER{\prt^m \prtb^n \bar{\zeta}}  \\[2mm]\nonumber
 - \prt^{m} \prtb^n (\prtb \rho^{z^n} \zbz- \mu\prt\rho^{z^n} \zbz
+n\prt\mu\rho^{z^n} \zbz)
 \DER{\prt^m \prtb^n \beta_{\bz}^{z^n}}\\[2mm]\nonumber
\left. + \prt^m \prtb^{n}(\prt\bar{\rho}^{\bz^n} \zbz-
\mub\prtb\bar{\rho}^{\bz^n} \zbz
+n\prtb\mub\bar{\rho}^{\bz^n} \zbz) \DER{\prt^m \prtb^n
\bar{\beta}_{z}^
{\bz^n}}\right)\\[2mm]\nonumber
\lbl{deltatilde1}
\nonumber
\ena

The previous operator being nilpotent, we filter it with the
counting operator
of the  fields :

\bea
\tld{\delta}(0)_0=\sum_{m,n \geq 0}   \left(
\prt^m \prtb^n \biggl(-2c_0(\prt \prtb )\phi_{00}\right)
 \DER{\prt^m \prtb^n \gamma_{1,1}}\\[2mm]\nonumber
+ \prt^m \prtb^n \biggl(-
\prtb\eta\zbz \biggr) \DER{\prt^m \prtb^n \zeta}  \\[2mm]\nonumber
+ \prt^m \prtb^n \biggl(-
\prt\bar{\eta}\zbz
 \biggr) \DER{\prt^m \prtb^n \bar{\zeta}}  \\[2mm]\nonumber
 +\prt^{m} \prtb^n (\prtb \rho^{z^n} \zbz)
 \DER{\prt^m \prtb^n \beta_{\bz}^{z^n}} \\[2mm]\nonumber
+ \prt^m \prtb^{n}(\prt\bar{\rho}^{\bz^n} \zbz) \DER{\prt^m \prtb^n
\bar{\beta}_{z}^
{\bz^n}}\biggr)\\[2mm]
\ena

So we can calculate the Laplacian:
\bea
\left\{\tld{\delta}(0)_0,\tld{\delta}(0)^{\dag}_0\right\}
\ena
getting:
\bea
\left\{\tld{\delta}(0)_0,\tld{\delta}(0)^{\dag}_0\right\}
=\sum_{m,n \geq 0}   \left(
\prt^m \prtb^n \biggl(2c_0(\prt \prtb )\phi_{00}\zbz\right)
\DER{\prt^m \prtb^n 2c_0(\prt \prtb )\phi_{00}}\\[2mm]\nonumber
+\prt^m \prtb^n \gamma_{1,1}
 \DER{\prt^m \prtb^n \gamma_{1,1}}
+ \prt^m \prtb^n \biggl(
\prtb\eta\zbz \biggr) \DER{\prt^m \prtb^n
\prtb\eta }  \\[2mm]\nonumber
+\prt^m \prtb^n \zeta\DER{\prt^m \prtb^n \zeta}
+ \prt^m \prtb^n \biggl(\prt\bar{\eta}\zbz \biggr) \DER{\prt^m
\prtb^n
\prt\bar{\eta}}  \\[2mm]\nonumber
+{\prt^m \prtb^n \bar{\zeta}}\DER{\prt^m \prtb^n \bar{\zeta}}
 +\prt^{m} \prtb^n (\prtb \rho^{z^n} \zbz)
 \DER{\prt^{m} \prtb^n (\prtb \rho^{z^n})} \\[2mm]\nonumber
 +\prt^m \prtb^n \beta_{\bz}^{z^n} \zbz)
 \DER{\prt^m \prtb^n \beta_{\bz}^{z^n}} \\[2mm]\nonumber
+ \prt^m \prtb^{n}(\prt\bar{\rho}^{\bz^n} \zbz) \DER{\prt^m
\prtb^{n}(
\prt\bar{\rho}^{\bz^n} )}
+ \prt^m \prtb^n \bar{\beta}_{z}^
{\bz^n} \zbz \DER{\prt^m \prtb^n \bar{\beta}_{z}^
{\bz^n}}\biggr)
\\[2mm]
\lbl{lapla1}
\ena

The solution of the system \rf{iso} will be into the kernel of the
previous
operator, so we can decompose \rf{lapla1} into a sum of positive
terms of moduli
(with respect of our definition of adjoint); and each of them will
be
identically zero.

It is matter of calculation to derive that the cohomological
space does not depend on
$${\beta}_{z}^{z^n} \zbz, \bar{{\beta}_{\bz}^{z^n}
\zbz},\gamma_{1,1}\zbz,
\zeta\zbz,\bar{\zeta}\zbz$$,
$$\prt \prtb\phi_{00}\zbz$$
$$\prtb\rho^n\zbz, \prt\bar{\rho}^n\zbz$$
and their derivatives.

We remark that the $\rho^{z^n} \zbz$ dependence of \rf{deltatilde1}
will come from the inhomogeneous term of the $\beta$ BRS
transformation, that is we
have to require:
$ (\prtb \rho^{z^n} \zbz
- \mu\prt\rho^{z^n} \zbz +n\prt\mu\rho^{z^n} \zbz) \neq 0 $.

The next filtration will provide

\bea
\tld{\delta}(0)_1=\sum_{m,n \geq 0}   \left(
\prt^m \prtb^n \biggl(-c_0{( - \mub
\prtb-\prtb\mub)(\prtb)\phi_{00}
-\prt (\mu \prt)\phi_{00}}
\right)
 \DER{\prt^m \prtb^n \gamma_{1,1}}\\[2mm]\nonumber
\prt^m \prtb^n \biggl(-c_0{(  \prt\phi_{00} \prt\phi_{00}) }
\biggr)
 \DER{\prt^m \prtb^n \eta}\\[2mm]\nonumber
\prt^m \prtb^n \biggl(-c_0{(  \prtb\phi_{00} \prtb\phi_{00} }
\biggr)
 \DER{\prt^m \prtb^n \bar{\eta}}\\[2mm]\nonumber
+ \prt^m \prtb^n \biggl(-\eta\zbz\prt\mu\zbz-\prt(\mu\zbz\eta\zbz)
\\[2mm]\nonumber
-\bar{\eta}\zbz\prt\mub\zbz
 \biggr) \DER{\prt^m \prtb^n \zeta}  \\[2mm]\nonumber
+ \prt^m \prtb^n
\biggl(-\bar{\eta}\zbz\prtb\mub\zbz-\prtb(\mub\zbz\bar{\eta}
\zbz)
\\[2mm]\nonumber
-{\eta}\zbz\prtb\mu\zbz)
 \biggr) \DER{\prt^m \prtb^n \bar{\zeta}}  \\[2mm]\nonumber
- \prt^{m} \prtb^n (- \mu\prt\rho^{z^n} \zbz
+n\prt\mu\rho^{z^n} \zbz)
 \DER{\prt^m \prtb^n \beta_{\bz}^{z^n}}\\[2mm]\nonumber
+ \prt^m \prtb^{n}(- \mub\prtb\bar{\rho}^{\bz^n} \zbz
+n\prtb\mub\bar{\rho}^{\bz^n} \zbz) \DER{\prt^m \prtb^n
\bar{\beta}_{z}^
{\bz^n}}\biggr)\\[2mm]
\nonumber
\ena
whose Laplacian zero modes, in the space of those of \rf{lapla1}
will
eliminate, after standard calculations, the dependence of the
cohomology space from $\eta$,$\bar{\eta}$,  their derivatives,
and from
$(\prt\prt) \phi_{00} \zbz$, $(\prtb\prtb) \phi_{00} \zbz$,
$\rho^n, \bar{\rho}^n, n\neq 0$
and their derivatives,

We have so shown that the cohomology space does not depend on the
negative
charged external fields, except  the terms depending from
underivated
$\rho^n, \bar{\rho}^n, n= 0$

Having so eliminated the $\Phi.\Pi.$ negative charged fields
content, the B.R.S.
operator \rf{delta00} strictly recall the one discussed in
\cite{theone};
so the next calculations will be similar as the ones in this
references.

The second filtration will led to the operator:
\eq
\tld{\delta}(1) \equiv \prt c\zbz S + \prt \bc\zbz T
+ \prtb c\zbz \bar{T} + \prtb \bc\zbz \bar{S} + R(1)\ .
\lbl{filtra1}
\en
where:
\nnbea
S^0 = \sum_{m,n \geq 0} \left\{ m \left(
\prt^m \prtb^n \phi_{0,0} \zbz \DER{\prt^m \prtb^n \phi_{0,0}}
+ \prt^m \prtb^n \mu\zbz \DER{\prt^m \prtb^n \mu} \right. \right.
\\[2mm]
\left. + \prt^m \prtb^n \mub \zbz \DER{\prt^m \prtb^n \mub}
+ \prt^m \prtb^n \lambda \zbz \DER{\prt^m \prtb^n \lambda}
+ \prt^m \prtb^n \lambdab \zbz \DER{\prt^m \prtb^n \lambdab}
\right) \\[2mm]
+ \prt^m \prtb^n \lambda \zbz \DER{\prt^m \prtb^n \lambda} \\[2mm]
\left. - \prt^m \prtb^n \mu\zbz \DER{\prt^m \prtb^n \mu}
+ \prt^m \prtb^n \mub \zbz \DER{\prt^m \prtb^n \mub} \right\}\ ,
\nnena

\nnbea
S^1 = \sum_{m+n \geq 1} \left\{ m \left(
\prt^m \prtb^n c\zbz \DER{\prt^m \prtb^n c}
+ \prt^m \prtb^n \bc\zbz \DER{\prt^m \prtb^n \bc} \right) \right.
\\[2mm]
\left. - \prt^m \prtb^n c\zbz \DER{\prt^m \prtb^n c} \right\} \ .
\nnena

Note that $S$ is nothing else that the "little z" indices counting
operator
$N_{z}(\downarrow)-N_{z}(\uparrow)$;that is:

\eq
S =N_{\prt} + N_{\lambda} + N_{\mub} - N_{\mu} - N_{c} \,
\equiv N_{z}(\downarrow)-N_{z}(\uparrow)\
\lbl{count}
\en

Moreover $S^{\dag}=S$; similarly:

\eq
\bar{S} = N_{\prtb} + N_{\lambdab} + N_{\mu} -
N_{\mub} - N_{\bc} \equiv N_{\bz}(\downarrow)-N_{\bz}(\uparrow)\ ,
\lbl{countb}
\en
and $\bar{S}^{\dag}=\bar{S}$.

Furthermore:

\nnbea \leqa{
T^0 = \sum_{m+n \geq 1} \left\{ m \left(
\prt^{m-1} \prtb^{n+1} \phi_{0,0} \zbz \DER{\prt^m \prtb^n
\phi_{0,0}}
\right. \right. }\\[2mm]
&& + \prt^{m-1} \prtb^{n+1} \mu\zbz \DER{\prt^m \prtb^n \mu}
+ \prt^{m-1} \prtb^{n+1} \mub \zbz \DER{\prt^m \prtb^n \mub}
\\[2mm]
&&\left. + \prt^{m-1}  \prtb^{n+1} \lambda \zbz \DER{\prt^m \prtb^n
\lambda}
+ \prt^{m-1} \prtb^{n+1} \lambdab \zbz \DER{\prt^m \prtb^n
\lambdab} \right)
\\[2mm]
&&+ \prt^m \prtb^n (\lambda \mu) \zbz \DER{\prt^m \prtb^n \lambda}
\\[2mm]
&&\left. - \prt^m \prtb^n \mu^2 \zbz \DER{\prt^m \prtb^n \mu}
\right\}
+ \DER{\mub}\ ,
\nnena
and
\nnbea
T^1 = \sum_{m+n \geq 2} \left\{ m \left(
\prt^{m-1} \prtb^{n+1} c\zbz \DER{\prt^m \prtb^n c}
+ \prt^{m-1} \prtb^{n+1} \bc\zbz \DER{\prt^m \prtb^n \bc} \right)
\right.
\\[2mm]
\left. - \prt^m \prtb^n c\zbz \DER{\prt^m \prtb^n \bc} \right\}
+ \prtb c \DER{\prt c} \equiv T^1_2 + \prtb c \DER{\prt c} \ .
\nnena

\eq
R(1) = \sum_{k+l \geq 2 } \grt{(} \prt^k\prtb^l c \zbz R_{kl}
+ \prt^k\prtb^l \bc \zbz \bar{R}_{kl}\zbz \grt{)} \ ,
\lbl{R}
\en

\nnbea \leqa{
R_{kl} \zbz = \sum_{m+n > k+l;m\geq k;n\geq l}
\left( \begin{array}{c} m \\ k \end{array} \right)
\left( \begin{array}{c} n \\ l \end{array} \right) \times } \\[2mm]
&& \left( \prt^{m-k+1} \prtb^{n-l} c \zbz \DER{\prt^m \prtb^n c}
+ \prt^{m-k+1} \prtb^{n-l} \bc \zbz \DER{\prt^m \prtb^n \bc}
\right)\ ,
\nnena
and $\bar{R}_{kl} = R_{k+1,l-1}$.

The Spectral Sequence analysys can be applied to $\tld{\delta}(1)$,
since:

\eq
\tld{\delta}(1)^2 = 0\ .
\lbl{crux}
\en

Filtering now with:

\eq
\nu' = 1 + \prt c \DER{\prt c} + \prtb \bc \DER{\prtb \bc}
+ 2\left( \prtb c \DER{\prtb c} + \prt \bc \DER{\prt \bc} \right) \
,
\lbl{nu'}
\en
we have the finite filtration:

\eq
\grt{[} \nu' , \tld{\delta}(1) \grt{]} = \sum_{n=1}^4 n \,
\tld{\delta}'(n)\ ,
\lbl{filtra'}
\en
with
\bea
\tld{\delta}'(1) &=& R(1) \nn[1mm]
\tld{\delta}'(2) &=& \prt c\zbz S + \prtb \bc\zbz \bar{S} \nn
&&\mbox{}\lbl{filtra2}\\[-3mm]
\tld{\delta}'(3) &=& \prt \bc\zbz T^1_2 + \prtb c\zbz \bar{T}^1_2
\nn[1mm]
\tld{\delta}'(4) &=& \prt\bc\zbz \prtb c\zbz
\left( \DER{\prt c} - \DER{\prtb \bc} \right) \nn
\ena
and we have to solve:

\eq
\left\{ \begin{array}{l}
\tld{\delta}'(n) \tld{\Delta} \zbz=0 \\[2mm]
\tld{\delta}'^{\dag}(n) \tld{\Delta} \zbz=0
\end{array} \right. \hs{5} \mbox{\normalsize for}\ n=1,\dots,4.
\lbl{iso'}
\en
which is equivalent to:
\eq
<\! \tld{\Delta}\zbz| \grt{\{} \tld{\delta}'^{\dag}(n) ,
\tld{\delta}'('n) \grt{\}}
|\tld{\Delta}\zbz\!> = || \tld{\delta}(n) \tld{\Delta}\zbz||^2
+ || \tld{\delta}'^{\dag}(n) \tld{\Delta}\zbz||^2 = 0 \ ,
\lbl{lapla}
\en
where the scalar product  is the one induced through our definition
of the
adjoint.

First for n=2 we get:

\eq
\begin{array}{ccl}
<\! \tld{\Delta}\zbz| \grt{\{} \tld{\delta}'^{\dag}(2) ,
\tld{\delta}'(2) \grt{\}}
|\tld{\Delta}\zbz\!> &=& || \tld{\delta}'(2) \tld{\Delta}\zbz||^2
+ || \tld{\delta}'^{\dag}(2) \tld{\Delta}\zbz||^2 \\[2mm]
&=& || S \tld{\Delta}\zbz||^2 + || \bar{S} \tld{\Delta}\zbz||^2 = 0
\ ,
\end{array}
\lbl{n2}
\en
which is solved by:
\eq
\left\{ \begin{array}{l}
S \tld{\Delta}\zbz = (N_{z}(\downarrow)-N_{z}(\uparrow))
\tld{\Delta}\zbz = 0\\[2mm]
\bar{S} \tld{\Delta}\zbz = (N_{\bz}(\downarrow)-N_{\bz}(\uparrow))
\tld{\Delta}\zbz = 0
\end{array} \right.\ ,
\lbl{soln2}
\en
where $\tld{\Delta}\zbz$ do not contain underivated
$c\zbz$,$\bc\zbz$ ghosts,
which are the only fields display an upper "little "index content.

So \rf{soln2} implies that the "little" indices indices have to be
saturated.
Now this constraint imposed by the cohomology, combined with the
fact that
underivated ghosts are absent; it is easy to realize that
space-time derivatives
of $c\zbz$ and $\bc\zbz$ of order greater than two are absent,
since in this
case any "little" indices saturation can be insured.

So the n=1 condition:

\ceq
\left\{ \begin{array}{l}
\tld{\delta}'(1) \tld{\Delta} \zbz=0 \\[2mm]
\tld{\delta}'^{\dag}(1) \tld{\Delta} \zbz=0
\end{array} \right.
\cen
is easily verified.

Proceeding further, for n=3 we have:

\nnbea \leqa{
<\! \tld{\Delta}\zbz| \grt{\{} \tld{\delta}'^{\dag}(3) ,
\tld{\delta}'(3) \grt{\}}
|\tld{\Delta}\zbz\!>\ =\ || \tld{\delta}'(3) \tld{\Delta}\zbz||^2
+ || \tld{\delta}'^{\dag}(3) \tld{\Delta}\zbz||^2 }\\[2mm]
&=& || T^0 \tld{\Delta}\zbz||^2 + || \bar{T}^0 \tld{\Delta}\zbz||^2
\\[2mm]
&+& <\!\tld{\Delta}\zbz| \prt \bc\zbz
[T,T^{\dag}]\DER{\prt\bc}|\tld{\Delta}\zbz\!>
\\[2mm]
&+& <\! \tld{\Delta}\zbz| \prtb c\zbz [\bar{T},\bar{T}^{\dag}]
\DER{\prtb c}|\tld{\Delta}\zbz\!>\\[2mm]
&+& <\!\tld{\Delta}\zbz|\prtb c\zbz
[\bar{T},T^{\dag}]\DER{\prt\bc}|\tld{\Delta}\zbz\!> \\[2mm]
&+& <\!\tld{\Delta}\zbz|\prt\bc\zbz [T,\bar{T}^{\dag}]\DER{\prtb
c}|\tld{\Delta}\zbz\!>
\\[2mm]
\nnena
\nnbea \leqa{}
&=& || T^0 \tld{\Delta}\zbz||^2 + || \bar{T}^0 \tld{\Delta}\zbz||^2
+ || \DER{\prtb c}\tld{\Delta}\zbz||^2 \\[2mm]
&+& || \DER{\prt\bc} \tld{\Delta}\zbz||^2
+ ||L(\dots) \DER{\prtb c} \tld{\Delta}\zbz ||^2
+ || M(\dots) \DER{\prt\bc} \tld{\Delta}\zbz ||^2 = 0 \ ,
\nnena
where $L(\dots)$, $M(\dots)$ are complicated functions.

The posivity of the metric will imply:

\eq
\DER{\prtb c} \tld{\Delta}\zbz = \DER{\prt\bc} \tld{\Delta}\zbz = 0
\ .
\lbl{n3}
\en
this result tell us that the cohomology space does not contain
$\prtb c\zbz$,
$\prt\bc\zbz$ monomials.

At the end for n=4 we get:

\nnbea
<\! \tld{\Delta}\zbz| \grt{\{} \tld{\delta}'^{\dag}(4) ,
\tld{\delta}'(4) \grt{\}}
|\tld{\Delta}\zbz\!> &=&
|| \grt{(} \DER{\prt c} - \DER{\prtb\bc} \grt{)} \tld{\Delta}\zbz
||^2 \\[2mm]
&+& || \DER{\prtb c} \DER{\prt \bc} \tld{\Delta}\zbz ||^2 \\[2mm]
&-& || \DER{\prtb c} \grt{(} \DER{\prt c} -
\DER{\prtb\bc} \grt{)} \tld{\Delta}\zbz ||^2 \\[2mm]
&-& || \DER{\prt \bc} \grt{(} \DER{\prt c} -
\DER{\prtb\bc} \grt{)} \tld{\Delta}\zbz ||^2 = 0 \ .
\nnena
which, with the previous results \rf{n3} gives the information that

$\tld{\Delta}\zbz$ does not contain combination of monomials $\prt
c\zbz-
\prtb\bc\zbz$

Collecting together the results, the $\Phi, \Pi$ charged sector
will only
contain elements of the type:

\eq
\tld{\Delta}\zbz = \grt{(} \prt c\zbz + \prtb \bc\zbz \grt{)}
\Delta_0 \zbz \ ,
\lbl{sol}
\en

But:

\eq
\prt c\zbz + \prtb \bc\zbz = \tld{\delta} \ln \grt{(} \lambda \zbz
\lambdab \zbz
(1-\mmb \zbz) \grt{)} \ .
\lbl{nonlo}
\en

So if we enlarge the basis of the cohomology by introducing the non
local
(in $\mu$!) functions:
\ceq
\{ \ln \lambda , \ln \lambdab \} \ ,
\cen
the $\tld{\delta}$-cohomology will be empty.

On the other hand in the $\Phi, \Pi$ uncharged  space we have to
verify:

\eq \left\{ \begin{array}{l}
S^0 \Delta_0 \zbz = (N_{z}(\downarrow)-N_{z}(\uparrow)) \Delta_0
\zbz = 0\\[2mm]
\bar{S}^0\Delta_0\zbz=(N_{\bz}(\downarrow)-N_{\bz}(\uparrow))\Delta_0\zbz
= 0\\[2mm]
T^0 \Delta_0 \zbz = 0 \\[2mm]
\bar{T}^0 \Delta_0 \zbz = 0
\end{array} \right. \ .
\lbl{matt}
\en

So we have the following theorem:

 \vskip 0.5cm

{\bf Theorem}
{\em
 The ghost sectors ($\Phi$-$\Pi$ charge sectors) of the
$\tld{\delta}$-cohomology in the space of analytic functions of the
fields,
where the fields $\lambda$ and $\lambdab$  satisfy
the equation \rf{intfac}, and completed with the terms
$\{\ln\lambda,\ln\lambdab\}$,seen as independent fields
depend only on terms containing underivated
$\rho$ which multiply zero ghost sector elements of the cohomology.

The zero ghost sector is, on the other hand, non-trivial only in
the part
which contains matter fields. Its elements will contain
no free $z$ and $\bz$ indices, i.e. they are "scalar-like"
quantities with
respect to "little indices" but can hold the tensorial content with
respect
the "big indices" $Z$ and $\bZ$.
A generic element of
this space will be a $(h,\bar{h})$-conformal quantity of the form
\[ [f( \prt^m_Z \prt^n_{\bZ} \phi_{0,0} \ZBZ )]_{h,\bar{h}},\]
where  $f$ is an analytic function, (polynomial).
}

\end{document}